\documentclass[apsmath, apssymb]{revtex4} 

\usepackage{graphicx}
\begin{document} 
\title{ Self-tuning to the Hopf bifurcation in fluctuating systems}
\author{ J.Balakrishnan\thanks{E-mail: janaki@mis.mpg,de}\\ 
Max Planck Institute for Mathematics in the Sciences, \\
Inselstrasse 22, D-04103 Leipzig, ~Germany.}
\vspace{2in}
\vspace{2.5cm} 
\begin{abstract} 
\vspace{3.5cm} 
\vspace{2.5cm}
\noindent {\large {\bf Abstract}} 

\noindent The problem of self-tuning a system to the Hopf bifurcation in 
the presence of noise and periodic external forcing is discussed. 
We find that the response of the system has a non-monotonic dependence 
on the noise-strength, and displays an amplified response which is more 
pronounced for weaker signals. The observed effect is to be distinguished 
from stochastic resonance. 
For the feedback we have studied, the unforced self-tuned Hopf oscillator 
in the presence of fluctuations exhibits sharp peaks in its spectrum. 
The implications of our general results are briefly discussed in the context of 
sound detection by the inner ear.\\ 

P.A.C.S. numbers ~~: 05.40.-a, ~02.50.Ey, ~05.10.Gg, ~05.45.-a, ~05.40.Ca, ~87.10.+e  \\ 

{\bf Published in}: {\em J.Phys.A} {\bf 38}, 1627 (2005).\\ 

\end{abstract} 
\maketitle 
\vspace{1in}
\hrule
\vspace{0.1cm}
{\it Preprint}~~~~~~~~~~~ ~~~~~~~~~~~ ~~~~~~~~~~~ ~~~~~~~~~ ~~~~~ ~~~~~~ 
~~~~~~~~~~~ ~~~~~~~~~~~ ~~~~~~~~~  ~~~~~~~ {\it  April 2004} 
\newpage 
\subsection*{1. ~Introduction}

\noindent An extensive literature exists on the role of noise in dynamical systems 
(see, for example, [1]). Critical systems subject to external forces and 
noise are especially interesting and manifest novel effects in diverse situations
--- in pattern formation in chemical reaction-diffusion systems, 
fluid mechanics and turbulence, electrical engineering, lasers and biological 
systems. 
In several of these situations, it is desirable to control the operating point 
at an optimally convenient level to get the desired dynamics. The deceptively 
simple act of balancing a stick is an example of a situation where one has to 
continually self-tune the system to the stable point [2].\\ 

\noindent In this paper we study the problem of self-tuning a dynamical 
oscillatory system subject to fluctuations from the environment and periodic 
forcing, to the Hopf bifurcation. 
This exercise is motivated by examples from biological systems, such as 
the spiking of neurons and especially by the problem of hearing and sound 
detection by the inner ear.\\ 
We use a general feedback equation in which the control parameter is partly 
generated by the dynamics of the system.  Since the dynamical system we study 
is very general and generic, one can use our results for specific cases, such 
as for the system of the hair cell, which is the mechanoreceptor cell in the 
inner ear, responsible for detecting sound.\\  

\noindent The paper is organised as follows. Section 2 consists of two 
subsections: in the first, we give a brief discussion of a biological 
example of sound detection by the ear to motivate our study. The second 
subsection is a small discussion of a generic Hopf oscillator and introduces 
the form of the feedback we have considered in the paper. 
Section 3 discusses the dynamics of a resonantly forced self-tuned generic 
Hopf oscillator 
in the presence of additive white noise. A solution of the Fokker-Planck equation 
in a particular frequency regime is given in Section 4, and some ideas and 
speculations  
presented on the possible applications of our results in the context of 
signal detection and hearing. \\ 
We obtain several new and interesting results --- the system exhibits a 
highly amplified response which has a non-monotonic dependence on the 
noise-strength and which is more pronounced for weaker signals. In the 
biological context, this feature, of sharper sensitivity to weaker signals, 
is in accord with observations made of the inner ear [5]. 
The underlying mechanism seems to be some kind of a stochastic ``filtering'' 
effect and is unrelated to stochastic resonance. \\ 
The distance of the control parameter from its critical value at the 
bifurcation for the system is computed as a measure of self-tuning. 
In the weak-noise limit, for the unforced self-tuned noisy Hopf oscillator, 
we find that for each noise-strength it is periodically modulated. Also, at 
certain frequencies, the oscillator makes sudden large departures from the 
bifurcation point. 
For the forced oscillator this distance increases with increasing noise 
strength in the weak noise limit, first very rapidly, and then gradually 
plateaus off. \\ 
We also find that the self-tuned unforced Hopf oscillator subject to 
fluctuations would exhibit delta peaks in its spectrum. 
In the context of hearing, we suggest that this could be the origin of the 
sharp peaks observed in the spontaneous otoacoustic emission spectra. \\
Finally in the Conclusion, we summarize the main results obtained in the paper. 
Some of the lengthy derivations have been presented in two Appendices. 
  
\subsection*{2. ~Spontaneous oscillations \& Hopf bifurcation }

\noindent {\bf A. ~An example from biology: Sound detection by the ear}  

\noindent To motivate the study carried out in the paper, we provide here 
a brief outline of an interesting example found in a recent stream of 
research in biology which models the mechanotransducer cells in the ear
 which detect sound, as Hopf oscillators arranged along the cochlea. \\  
The sensory receptors of the inner ear are the hair cells. 
In the cochlea these are responsible for our sensitivity to sound, and 
in the vestibule, to our sense of balance. Mechanical stimuli are 
received by hair bundles --- assemblies of 20-300 cylindrical processes  
called the stereocilia (each consisting of an actin cytoskeleton ensheathed 
by a plasma membrane) located at the apex of the hair cell, and at their 
basolateral surfaces the hair cells make synapses with axons. 
Mechanical stimuli result in the rapid opening of ion channels located 
in the hair bundles, letting in an influx of current which results in 
electrical signals being conveyed to the nerve cells (see [3-6] for example, 
and references therein). \\ 
As the ear is able to respond to a wide range of stimuli, varying 
by several orders of magnitude [5], its responsiveness must therefore be 
nonlinear. 
At the lower end, the mammalian cochlea can respond to sound-driven 
vibrations of atomic dimensions, even as low as about $\pm 0.3$nm.  
In order to amplify such low intensity stimuli, it is believed that 
the ears have developed active force-generating dynamical processes [7-13]. 
The presence of such elements would also help to understand the recorded 
otoacoustic emissions (both spontaneous \& evoked) from the hearing 
organ of all organisms [8,9]. \\ 
Active hair bundle motion can account for many properties of the observed 
active processes, such as amplification, distortion tone production, 
spontaneous oscillations, otoacoustic emissions, high frequency selectivity 
and sharp sensitivity. A recent stream of research in the biophysics literature 
[10-14] identifies these features with signatures of a dynamical system 
operating near a Hopf bifurcation. The system acts in the immediate 
vicinity of the bifurcation as a nonlinear amplifier for sinusoidal 
stimuli close to the characteristic frequency. As the control parameter 
is varied, the system changes behaviour from a quiescent state and 
exhibits self-sustained oscillations. Effective amplification is achieved 
by tuning the system to operate close to the bifurcation. The concept of 
a ``self-tuned Hopf bifurcation'' was invoked in [11], emphasizing that a 
self-regulatory mechanism exists which enables the hair cell to operate in 
the proximity of the critical point, whatever its characteristic frequency. 
This work did not, however, discuss the mechanism of self-tuning. 
Using two adaptation mechanisms, it was demonstrated in [13] that the hair 
cell can produce ``self-tuned critical oscillations''. 

\noindent In [15], a mathematical study of feedback tuning to the bifurcation 
was discussed for a neural integrator and for the hair cell in the absence 
of fluctuations. Their feedback for the system of the hair cell differs 
from that in [13], and from what is normally found in the 
biophysically-oriented literature. \\ 

\noindent In their very interesting work, Jaramillo \& Wiesenfeld [16], 
inspired by the seminal experiments reported in [17] and the work in [18-23], 
proposed that 
Brownian motion of the hair bundle enhances the sensitivity of mechanoelectrical 
transduction by the inner hair cells, overcoming the huge viscous drag force 
of the fluid in which they are immersed and amplifying limitingly low-amplitude 
stimuli through stochastic resonance [22-26]. Their mechanism was based on the 
well-established 
gating-spring model [3] of mechanotransduction, and they provided as an example, 
the two-state model, the open and closed states of the transduction channels in 
the stereocilia representing the two stable states within the framework of the 
bistable models of stochastic resonance [22,23]. \\ 

\noindent While their conjecture is highly attractive, it is still interesting 
to see how  an active system like a single hair cell as a unit by itself 
responds to fluctuations and stimuli from outside, since its dynamics is 
controlled by feedback mechanisms -- the calcium dynamics through the 
transduction channels being one such example.  
The hypothesis that the activity of the hair-cell is due to the operation 
of its kinematical constituents at the edge of a Hopf instability, was first 
worked out explicitly in [10]. If one was to accept this line of thinking, 
it would be of interest to see how well a generic Hopf oscillator, whose 
operating point is controlled through a feedback mechanism, would act as a detector. 
This is our motivation for studying self-tuning in a forced Hopf oscillator 
subject to noise.\\ 
The form of the feedback we have studied, which is partly 
generated by the dynamics of the system is motivated by the biological 
system of the hair bundle. Studying a generic system enables  
us to capture the essential behaviour of the simplified system dynamics 
and to offer possible explanations of various phenomena without having to 
deal with the vastly cluttered and often messy equations in complex biological 
systems. Thus as with all models of biological systems, substitution of actual 
parameters coming from the mechanoelectrical and chemical properties of a 
hair cell in the results we have obtained for a generic system could be expected 
to give better and more realistic estimates of how well a Hopf oscillator models 
the real biological hair cell.  \\

\noindent A model to describe active hair bundle motion has been presented in the 
past by several authors [4,11,13] in the following way. 
Let $x(t)$ denote the displacement of the hair bundle to a periodic external 
stimulus $F_{ext}$ of frequency $\Omega$ and $y(t)$ denote the force generated 
by the active component of the hair bundle arising from the motion of the 
channel motor complex along an actin filament in the stereocilium. The linear 
behaviour of the hair bundle displacement may be described by the equation: 
\begin{eqnarray} 
\lambda\frac{dx}{dt} &=& -kx + y + F_{ext}(t) \nonumber\\
\beta\frac{dy}{dt} &=& -y - {\bar k}x 
\end{eqnarray} 
where $\lambda$, $k$ and $\beta$ stand for the drag coefficient, stiffness of 
the hair bundle and relaxation time of the active process and ${\bar k}$ has 
dimensions of a spring constant. The second equation is obeyed by the force 
generated by the active process within the hair bundle. The model (eqn.(1)) 
is described in Ref.[4,6] as 
also in [11,13] and arises from the gating-spring model for regulation of a 
mechanoreceptive ion channel in the stereocilia, in which each channel exists 
within two conformations -- the open and the closed states, and opening of a 
channel shortens the associated gating spring and its tension, lowering the 
energy of that state in comparison with the closed state. \\ 

\noindent The nonlinear terms in the equations arise from this swing of the 
gating spring which depends upon the channel open probability which 
is a sigmoidal function of the hair bundle displacement. \\ 
A linear stability analysis on this system shows that the system exhibits a 
Hopf bifurcation when ~$\frac{1}{\beta} = -\frac{k}{\lambda}$. ~These coupled 
equations can be rewritten as a single equation for a complex variable ~$z$ 
\begin{equation}
z = x + i\frac{y}{k}  ~~.
\end{equation} 
It is possible to rewrite the resulting equation in ~$z$ in a canonical normal 
form through a nonlinear transformation of the variable. \\

\noindent {\bf B. ~Generic Hopf oscillator}  

\noindent In the presence of an external driving force  ~$F_{ext} = F_0\cos\Omega t $, 
the generic normal form for a system such as eqn.(1) near a Hopf bifurcation has the 
form [28-31]  
\begin{equation} 
\frac{dz}{dt} = A(\omega, C)z - B(\omega, C)|z|^2z + O(|z|^4z) + 
F_0 e^{i\alpha_0} e^{i\Omega t}{\bar z}^{s-1}   
\end{equation} 
where ~$A$ and ~$B$  are complex coefficients which depend upon the characteristic  
frequency ~$\omega$ of the Hopf oscillator and on the bifurcation parameter 
~$C$ of the system. The additional factor ~$e^{i\omega_0}$ arises from the 
transformation of the original equation to the normal form (eqn.(3)) and is 
independent of ~$z$ and ~$\bar z$ (for the hair bundle, ~$\alpha_0$ depends 
upon its mechanical parameters). \\
Equation (3) describes the system at  ~$s:m$ resonance ~($s, m$ being coprime 
integers),  ~$\Omega = \omega(\frac{s}{m} + \gamma)$ where $\gamma$ is the 
detuning parameter, 
and ~$F_0$ denotes the amplitude of forcing [29].\\ 
In this paper, we consider the situation in which the Hopf oscillator is in 1:1 
resonance with the driving frequency, and $\gamma =0$. \\  
The control parameter, $C(t)$, regulates the dynamics of the oscillating system 
and tunes it to operate very close to the dynamical Hopf instability. In this paper, 
we study a Hopf oscillator having a feedback of the form  
\begin{equation}
\frac{dC}{dt} = \Gamma(C, x(t)) 
\end{equation} 
where ~$\Gamma(C,x(t))$ is, in general, a nonlinear function of $C$, and one of the 
dynamical variables ~$x(t)$ of the system.  

In the case of the hair bundle, the 
control parameter may be taken as the concentration of the 
calcium entering the stereocilia, which regulates the opening and closing of ion 
channels present on them [13]. 
Here, the dynamics of the control parameter is intertwined with the dynamical 
motion and displacement of the hair bundle so that ~$C = C(t, x(t))$. Thus one 
could apply the results of our analysis for a generic Hopf oscillator 
having the feedback control (eqn.(4)) we have considered, to the system of 
the hair bundle.  

For a generic Hopf oscillator, a solution of eqn.(3) can be found in terms of 
its normal modes using Fourier expansions, using the method in [11]. 
We give the details of this procedure in Appendix-A, stating the result below. 
A Fourier expansion of the variable ~$x$ is made and it is assumed that the 
first Fourier mode ~$x_1$ is the dominant one near the bifurcation. 
In the absence of external forcing, this leads to the equation  
\begin{equation}
0  \approx \lambda_6(C_0)x_1 + \lambda_8(C_0){|x_1|}^2x_1 
\end{equation} 
where ~$\lambda_6(C_0)$ and $\lambda_8(C_0)$ are some functions of $C_0$.  
Spontaneous oscillations occur when 
\begin{equation}
{|x_1|}^2 \approx -\frac{\lambda_6(C_0)}{\lambda_8(C_0)}   ~~~~, 
\end{equation}
for nonzero $\lambda_6$, when the right hand side of eqn.(6) is positive. 
This can be rewritten in the form expressed in [11] as 
\begin{equation}
|x_1| \approx \Delta{\Big( \frac{C_c - C}{C_c} \Big)}^{1/2} ,
\end{equation}
where ~$\Delta$ represents a characteristic saturating value for the 
$x$ variable and $C_c$ is the critical value of the control 
parameter. The form expressed in eqn.(7), enables one to keep track of 
the distance of the control parameter from its critical value.\\

In this paper we do not restrict ourself to the specific problem of the 
hair bundle oscillations, but rather, we consider a generic Hopf oscillator 
system in which the control parameter is dependent upon the dynamics 
of the system (eqns.(4),(A-3)). \\

\subsection*{3. Self-tuned Hopf Oscillator in a fluctuating environment}

A general two-variable system in the absence of external forcing, containing a Hopf 
bifurcation and perturbed by an 
external additive noise ~$\xi(t)$ has the form [32]  
\begin{equation} 
d_t \left [ \begin{array}{ll}x\\ 
y  
\end{array} 
\right ] = \left[ \begin{array}{ll} \mu ~ -\omega\\ 
\omega ~~~~ \mu     
\end{array} 
\right]\left [ \begin{array}{ll}x\\ 
y  
\end{array} 
\right] + \left[ \begin{array}{ll} f_1x^2 + f_2xy + f_3y^2 + f_4x^3 + f_5x^2y 
+ f_6xy^2 + f_7y^3\\ 
g_1x^2 + g_2xy + g_3y^2 + g_4x^3 + g_5x^2y + g_6xy^2 + g_7y^3     
\end{array} 
\right]   
+ \left[ \begin{array}{ll}\sigma_1\\ 
\sigma_2   
\end{array} 
\right]\xi(t) + O(4)
\end{equation}
before reduction to the normal form. Here, ~$\omega$, the frequency of 
the oscillator and ~$\mu$, which unfolds the bifurcation, are the imaginary 
and real parts respectively of ~$A(\omega, C)$.  The coefficients 
~$f_i$ and ~$g_i$ ~($i = 1, \dots n$) are operators in general -- we consider 
systems which allow for periodic orbits, so the coefficients are such that the  
system cannot be rewritten in a gradient form.  

We start by observing that very close to the Hopf bifurcation, the 
coefficients $A$ and $B$ in eqn.(3) for the system with self-tuning must 
be understood as being Taylor expansions around the critical value 
$C_c$ of the control parameter.  
\begin{eqnarray}
A(\omega, C) &=& A(C_c) + A'(C_c)(C_0 - C_c) + \frac{1}{2}{(C_0 - C_c)}^2 
A''(C_c) + O(A''')\nonumber\\
B(\omega, C) &=& B(C_c) + B'(C_c)(C_0 - C_c) + \frac{1}{2}{(C_0 - C_c)}^2
B''(C_p) + O(B''')
\end{eqnarray} 

\noindent From eqn.(7) we see that for the system described by eqn.(3), spontaneous 
oscillations become possible when 
\begin{equation}
C_0 - C_c \approx \frac{C_c}{\Delta^2}{|x_1|}^2 
\end{equation}
We substitute this into eqn.(9) and use the resulting expressions in eqn.(3) to write 
down the following normal form equation for a generic Hopf oscillator with a 
feedback control eqn.(4), and perturbed by an external periodic forcing 
\begin{equation} 
\frac{dz}{dt} =  (A(C_c) + A'(C_c)\frac{C_c}{\Delta^2}{|x_1|}^2 + \dots )z - 
(B(C_c) + B'(C_c)\frac{C_c}{\Delta^2}{|x_1|}^2 + \dots )|z|^2z + O(|z|^4z) + 
F_0 e^{i\alpha_0} e^{i\Omega t}{\bar z}^{s-1}   
\end{equation}
As already mentioned, we study only the case of 1:1 resonance with $s=1$ in this 
paper.\\  
We introduce the coefficients  
\begin{eqnarray}
\beta &=& Re(A(C_c)) ; ~~~~~~ l = Re(B(C_c)) ; ~~~~~ \beta'(C_c) = Re(A'(C_c))
\nonumber\\ 
\omega &=&-Im(A(C_c)) ; ~~ d = -Im(B(C_c)) ; ~~~ \omega'(C_c) = -Im (A'(C_c))
\end{eqnarray} 
In order to study fluctuations in the system close to the bifurcation, we follow 
the method developed in [33] (see also [35]) for the case of the weak-noise limit. 
The system of equations (11) are then rewritten in the form  
\begin{equation}
\frac{dz}{dt} = f(z, \bar z, t) + \epsilon^{1/2}\xi(t) 
\end{equation}  
where ~$f(z, \bar z, t)$ includes all the deterministic terms on the right hand 
side of eqn.(11), ~$\xi(t) = \xi_1(t) + i\xi_2(t)$ denotes a (complex) white noise 
and we study the asymptotic 
behaviour of this stochastic process in the limit ~$\epsilon \rightarrow 0$. 
We define the noise correlations as  
\begin{eqnarray} 
\langle\xi_1(t)\xi_1(t')\rangle &=& Q_a\delta(t-t')\nonumber\\
\langle\xi_2(t)\xi_2(t')\rangle &=& Q_b\delta(t-t')\nonumber\\
\langle\xi_1(t)\xi_2(t')\rangle &=& Q_{ab}\delta(t-t')
\end{eqnarray}
We now write $z$ in polar coordinates  
\begin{equation} 
z(t) = r(t) e^{i\phi(t)}
\end{equation}   
We are interested in understanding how a Hopf oscillator having a feedback control 
behaves differently from one without self-tuning. To this end, we combine the 
Fourier expansion of the 
variable ~$x(t)$, its first Fourier mode being the dominant one, with 
the polar coordinate representation of ~$z$ in eqn.(15) to express ~$x_1$ as  
\begin{equation}
2 |x_1|\cos(\Omega t + \alpha) = r(t)\cos\phi(t)
\end{equation}
for nonzero ~$x$, to study the approach to criticality.  

\noindent Equation (11) can then be expressed as a set of coupled Langevin equations  
\begin{eqnarray}
{\dot r} &=& \beta r - \Big( l - \beta'\frac{C_c}{\Delta^2}\frac{\cos^2\phi(t)}
{4 \cos^2(\Omega t + \alpha)}\Big)r^3 + F_0\cos(\Omega t + \alpha_0 -\phi(t)) + 
O(r^5) + \alpha_1\xi_r \nonumber\\ 
{\dot \phi} &=& -\omega(C_c) - \Big(-d(C_c) + \omega'(C_c) \frac{C_c}{\Delta^2} 
\frac{\cos^2\phi(t)}{4 \cos^2(\Omega t + \alpha)} \Big )r^2 
+ F_0\frac{\sin(\Omega t + \alpha_0 -\phi(t))}{r} +O(r^4) + 
\frac{\alpha_2\xi_\theta}{r} \nonumber\\ 
\end{eqnarray}
where 
\begin{eqnarray}
\alpha_1\xi_r &=& (\cos\phi\xi_1 + \sin\phi\xi_2)\epsilon^{1/2} \nonumber\\
\alpha_2\xi_\theta &=& (\cos\phi\xi_2 - \sin\phi\xi_1)\epsilon^{1/2}
\end{eqnarray} 
Thus the noise now becomes process-dependent, as a result of the transformation 
to polar coordinates. \\
From here one can write the Fokker-Planck equation equivalent to eqn.(17). We find  
\begin{eqnarray} 
\frac{\partial P(r,\phi,t)}{\partial t} &=& - \frac{\partial}{\partial r}\Big[ 
\beta r - lr^3 + F_0\cos(\Omega t + \alpha_0 - \phi) + 
\beta'\frac{C_c}{\Delta^2}\frac{\cos^2\phi(t)}{4 \cos^2(\Omega t + \alpha)}r^3 
+ \frac{\epsilon}{2r}Q_{\phi\phi} \Big] P(r,\phi,t) \nonumber\\
&-&  \frac{\partial}{\partial \phi}\Big[ -\omega + d r^2 + 
\frac{F_0}{r}\sin(\Omega t + \alpha_0 - \phi) 
- \omega'\frac{C_c}{\Delta^2}\frac{\cos^2\phi(t)}{4 \cos^2(\Omega t + \alpha)}r^2 
- \frac{\epsilon}{r^2}Q_{r\phi} \Big] P(r,\phi,t) \nonumber\\
&+& \frac{\epsilon}{2}\Big[ \frac{\partial^2}{\partial r^2}Q_{rr} 
+ \frac{\partial^2}{\partial\phi^2}\frac{Q_{\phi\phi}}{r^2} 
+ 2 \frac{\partial^2}{\partial r\partial\phi}\frac{Q_{r\phi}}{r}\Big] P(r,\phi,t)
\end{eqnarray} 
where
\begin{eqnarray}
\frac{1}{2}Q_{rr} &=& Q_a\cos^2\phi + 2 Q_{ab}\sin\phi\cos\phi 
+ Q_b\sin^2\phi \nonumber\\
\frac{1}{2}Q_{r\phi} &=& - Q_a\cos\phi\sin\phi + Q_{ab}(\cos^2\phi -\sin^2\phi) 
+ Q_b\sin\phi\cos\phi \nonumber\\ 
\frac{1}{2}Q_{\phi\phi} &=& Q_a\sin^2\phi - 2 Q_{ab}\sin\phi\cos\phi + Q_b\cos^2\phi
\end{eqnarray} 
This is in accordance with Stratonovich calculus.\\
Let the deterministic part of eqn.(17) have a stable solution ~$(\bar r(t), \bar\phi(t))$. 
In this limit (i.e., with ~$\epsilon \rightarrow 0$), the probability density initially 
centred around 
 ~$(\bar r(t), \bar\phi(t))$  reduces for all times to 
\begin{equation}
\lim_{\epsilon \rightarrow 0} P(r,\phi,t) = \delta(r - \bar r(t))\delta(\phi 
- \bar\phi(t)) 
\end{equation}
$\bar r(t)$ and  ~$\bar\phi(t))$ obey 
\begin{eqnarray}
\frac{d\bar r}{dt} &=& \beta(C_c)\bar r 
- \Big(l-\beta'(C_c)\frac{C_c}{\Delta^2}\frac{\cos^2\bar\phi}{4\cos^2(\Omega t + \alpha)}
\Big)
{\bar r}^3  
+ F_0\cos(\Omega t + \alpha_0 -\bar\phi)\nonumber\\ 
\frac{d\bar \phi}{dt} &=& -\omega(C_c) - \Big(-d(C_c) + \omega'(C_c)\frac{C_c}{\Delta^2}
\frac{\cos^2\bar\phi}{4\cos^2(\Omega t + \alpha)}\Big){\bar r}^2 
+ \frac{F_0\sin(\Omega t 
+ \alpha_0 -\bar\phi)}{\bar r}
\end{eqnarray}
Thus, the equation for the radial variable is not separable from ~$\phi$, 
except for the special case ~$\phi = \Omega t + \alpha \pm 2n\pi , ~~(n = 0,1,2,\dots)$. 
For this case, one obtains in the absence of the external force,  
an orbitally stable periodic solution of circular form with amplitude
\begin{equation} 
{\bar r}_s = \pm {\Big[ \frac{\beta(C_c)}{\Big(l-\frac{\beta'(C_c)C_c}
{4\Delta^2}\Big)} 
\Big]}^{1/2} ~~~, ~~~~ {\rm for} ~~ 
C_c < \frac{4\Delta^2l}{\beta'}, ~~ \beta > 0 ,
\end{equation} 
when 
\begin{equation}
\bar\phi_s = \cos^{-1}{\Big( \frac{4d(C_c)\cos^2(\Omega t + \alpha)}{\omega'(C_c)}
\frac{\Delta^2}{C_c}\Big)}^{1/2} 
\end{equation} 
in the stationary state limit $t \rightarrow 0$. \\
Thus, when the feedback (eqns.(4),(A-1)) is switched on, the system  starts 
moving on a limit cycle of a larger radius. The sign and the magnitude of the 
critical value of the control parameter determines the sense of rotation of 
the limit cycle. \\

\noindent In order to see how fluctuations affect the regulatory role of the 
control parameter in bringing the system to operate in the close proximity of 
the bifurcation, we will define the distance ~$\delta C$~ of the control 
parameter at a given operating point from its critical value. The effect of 
fluctuations on  ~$\delta C$~ can be found by determining its noise average. 
In order to calculate noise averages, we must first determine the correct form 
of the probability distribution in the vicinity of the bifurcation. Hence the 
solution of the Fokker-Planck equation (19) must be found. \\
In order to do so, we use the singular perturbation technique as 
developed by Malek Mansour {\em et al.} [32] for studying fluctuations at 
the onset of a limit cycle. It was shown by these authors that the 
asymptotic properties of a stochastic process in the vicinity of the 
critical point could be studied by an appropriate scaling of the variables. 
In their procedure rescaled variables are introduced by expanding all 
quantities in terms of a single noise-smallness parameter ~$\epsilon$ 
(which is the inverse of the extensivity parameter $V$ which denotes 
the size of the system).  
Further they showed that the critical variable exhibits amplified non-Gaussian 
fluctuations on a slow time scale. 
  
Their arguments stem from their earlier result in the theory of stochastic 
processes that the stochastic and the macroscopic trajectories converge to 
a macroscopically steady state   
for all times, in the limit of weak noise ($\lim \epsilon \rightarrow 0)$, 
if the state is unique and globally stable. 
This is an extension of the well known theorem of Kurtz (see, for instance, 
reference [34]). 
For a two-variable system such as described by eqns.(8) and (13), the 
stochastic variables ~$x(t)$ 
and ~$y(t)$ can then be perturbatively expanded in powers of ~$\epsilon^{1-a}$, 
and ~$\epsilon^{1-b}$, ~($0 \le a,b <1$) around the 
deterministic steady state ~$\bar x(t)$, ~$\bar y(t)$. Scaled variables $u$ and $v$, 
can then be defined in terms of the deviations: ~$u = \epsilon^{a-1}(x-\bar x), 
~v = \epsilon^{b-1}(y-\bar y)$  and a probability density for the scaled 
variables helps to study the asymptotic properties of the process [33]. 
$a$ and $b$ are chosen so that the probability density remains normalizable 
in the weak noise limit.The values of $a$ and $b$ are further restricted 
to the range ~$\frac{1}{2} \le a,b <1$  by the requirements that the moments 
of the probability density for the scaled variables, if they exist 
for ~$\epsilon \ne 0$, remain finite in the limit ~$ \epsilon \rightarrow 0$, 
and that at least one of them be nonzero. We have ~$a = b = \frac{1}{2}$ when the 
initial distribution is Gaussian. The distribution does not depart from Gaussian 
behaviour if the fluctuations of the scaled variables do not diverge in the 
long time limit. 
The spectral properties of the matrix constructed from the linearization of 
the deterministic part of the general system presented in eqn.(8), which 
determines the stability of the macroscopic stationary state, also determines 
the asymptotic behaviour of the stochastic process. \\   

\noindent The description in terms of scaled variables enables one to 
determine the departure of the probability density for the stochastic process 
from its initial Gaussian behaviour for the macroscopic state [33]. Therefore, 
the scaling exponents provide, in a way, a measure of how much the 
fluctuations are amplified by the marginal stability of the state in 
the critical regime.  
We follow this procedure below for the stochastic system in eqns.(17),(19) 
whose deterministic part $(\bar r(t), \bar\phi(t))$  satisfies eqns.(21-24). \\ 

\noindent We consider in this paper only a soft transition leading to a limit cycle. 
Here also, as in [33], we can perform a linearization of the macroscopic  
state given by eqns.(21) and (22) around the stable state (eqns.(23,24)) 
and we are interested in the approach to the stationary state  
~$t \rightarrow \infty$. It was argued in [33], that since beyond the 
critical point, the radius ~${\bar r}(t)$ in eqn.(21) may evolve to a 
constant value ~${\bar r}_s$ even though periodicity in the phase variable 
is retained for all time, one can hence define a scaled variable ~$\rho$ as  
\begin{equation}
r = {\bar r}_s + \rho\epsilon^{1-b} ~~, ~~b<1 .
\end{equation} 
As we already discussed before, for ~$\epsilon = 0$, the probability 
distribution is a Dirac delta function (eqn.(21)) centred around the macroscopic 
state ~$(\bar r, \bar \phi)$. 
We wish to determine the form of the distribution for ~$\epsilon \ne 0$.\\ 
From eqns.(23) and (24), for   
~$\frac{\beta(C_c)}{\Big(l-\frac{\beta'(C_c)d(C_c)}{\omega'(C_c)}\Big)} < 0$~, 
~${\bar r}_s = 0$ ; and from eqn.(25)~ $b = 1/2$, which leads to 
a Gaussian probability distribution. \\
We now consider the case ~~$\frac{\beta(C_c)}{\Big(l-\frac{\beta'(C_c)d(C_c)}
{\omega'(C_c)}\Big)} \ge 0$ in which eqns.(22) admit an orbitally stable 
periodic solution as already discussed in eqns.(23) and (24). \\ 
In order to facilitate analysis of (19), we rewrite it in an autonomous form by 
extending the phase space through the transformation 
\begin{equation}
\lambda = \Omega t + \alpha
\end{equation} 
so that
\begin{equation}
\dot{\lambda} = \Omega 
\end{equation}
Close to the bifurcation, we scale ~$r$ as in eqn.(25) and the other quantities as 
\begin{eqnarray}
\beta &=& \tilde\beta\epsilon^{2c} \nonumber\\ 
{\bar r}_s \sim O(\epsilon^c) &=& \hat r_s\epsilon^c  ~~, ~~ \hat r_s \sim 
\frac{{\tilde\beta}^{1/2}}{l^{1/2}} 
~~~ ({\rm {because ~ of ~ eqn.(22)}} ) \nonumber\\ 
\phi &=& \phi \nonumber\\ 
F_0 = \hat{\hat F_0}{\tilde\beta}^{3/2}\epsilon^c &=& {\hat F_0}\epsilon^{3c} ~~~~~~~,
\end{eqnarray}
where ~$\tilde\beta > 0$ and is independent of ~$\epsilon$, ~$c>0$, 
and in the close vicinity of the bifurcation ~$c > 1-b$. 
In terms of the scaled variables, eqn.(19) can be rewritten as  
\begin{eqnarray}
\frac{\partial \wp(\rho,\phi,\lambda,t)}{\partial t} &=& -\frac{\partial}{\partial\rho}
\Big[ \frac{\epsilon^{2b-1}Q_{\phi\phi}}{2(\rho + {\hat r}_s\epsilon^{b+c-1})} 
+ \hat F_0\epsilon^{b+3c-1}\cos(\lambda -\phi +\alpha_0 -\alpha) 
+ \tilde\beta{\hat r}_s\epsilon^{b+3c-1} 
+ \tilde\beta\rho\epsilon^{2c} \nonumber\\ 
&-& \rho^3\Big( l - \beta'\frac{C_c}{\Delta^2}\frac{\cos^2\phi(t)}
{4 \cos^2\lambda}\Big)\epsilon^{2-2b} 
- 3\rho^2\Big( l - \beta'\frac{C_c}{\Delta^2}\frac{\cos^2\phi(t)}
{4 \cos^2\lambda}\Big){\hat r}_s\epsilon^{-b+c+1} \nonumber\\  
&-&3\rho\Big( l - \beta'\frac{C_c}{\Delta^2}\frac{\cos^2\phi(t)}
{4 \cos^2\lambda}\Big){{\hat r}_s}^2\epsilon^{2c} 
- \Big( l - \beta'\frac{C_c}{\Delta^2}\frac{\cos^2\phi(t)}
{4 \cos^2\lambda}\Big){{\hat r}_s}^3\epsilon^{b+3c-1} \Big] 
\wp(\rho,\phi,\lambda,t)\nonumber\\ 
&-&\frac{\partial}{\partial\phi}\Big[ -\omega + d \epsilon^{2-2b}\rho^2 
+ 2d \rho{\hat r}_s\epsilon^{-b+c+1} + d {{\hat r}_s}^2\epsilon^{2c} 
+ \frac{F_0\sin(\lambda-\phi +\alpha_0 -\alpha)
\epsilon^{b+3c-1}}{(\rho + {\hat r}_s\epsilon^{b+c-1})} 
\nonumber\\ 
&-&\omega'\frac{C_c}{\Delta^2}\frac{\cos^2\phi(t)}{4 \cos^2\lambda}
\Big( \rho^2\epsilon^{2-2b} + 2\rho{\hat r}_s\epsilon^{-b+c+1} 
+ {{\hat r}_s}^2\epsilon^{2c} \Big) - \frac{\epsilon^{2b-1}Q_{r\phi}}
{{(\rho + {\hat r}_s\epsilon^{b+c-1})}^2} \Big] \wp(\rho,\phi,\lambda,t) \nonumber\\ 
&+& \frac{1}{2}\Big[ \epsilon^{2b-1}\frac{\partial^2}{\partial\rho^2}Q_{rr} 
+ \epsilon^{2b-1}\frac{\partial^2}{\partial\phi^2}\frac{Q_{\phi\phi}}
{{(\rho + {\hat r}_s\epsilon^{b+c-1})}^2} + 2\epsilon^{2b-1}\frac{\partial^2}
{\partial\rho\partial\phi}\frac{Q_{r\phi}}{(\rho + {\hat r}_s\epsilon^{b+c-1})} 
\Big] \wp(\rho,\phi,\lambda,t) \nonumber\\ 
&-& \Omega\frac{\partial}{\partial\lambda} \wp(\rho,\phi,\lambda,t) 
\end{eqnarray} 
We must now determine the appropriate values of ~ $b$ and~ $c$ which would lead to 
a nontrivial form of the probability density of the scaled variables. 
Following the argument in [33], for ~$b\le 1/2$, ~to dominant order in ~$\epsilon$, 
one finds from eqn.(29) that  
\begin{equation}
\partial_t\langle\rho^2\rangle = \frac{Q_{rr}}{\epsilon^{1-2b}} > 0
\end{equation} 
which diverges as ~$\epsilon \rightarrow 0$, so that fluctuations in the long time 
limit of the radial variable are divergent. This leads to the conclusion that close 
to the bifurcation the long time regime must be described by ~$b>1/2$. \\ 
Using this result and integrating eqn.(29) with respect to ~$\phi$ we find that 
\begin{equation}
\partial_t \wp(\rho,\lambda,t) = O(\epsilon^{2b-1}, \epsilon^{2-2b}, \epsilon^{1-b+c}, 
\epsilon^{2c}, \epsilon^{b+3c-1}) \sim {\mathcal{Q}}(1)
\end{equation}
For a given ~$\rho$ if we denote the conditional probability of ~$\phi$ by  
~$\wp(\phi|\rho,\lambda,t)$, then from equations (29)-(31) we see that
\begin{equation}
\partial_t \wp(\phi|\rho,\lambda,t) + \Omega\frac{\partial}{\partial\lambda}
\wp(\phi|\rho,\lambda,t) = 
\omega\frac{\partial}{\partial\phi}\wp(\phi|\rho,\lambda,t) + {\mathcal{Q}}(1) . 
\end{equation}
This has the solution  
\begin{equation}
\lim_{\epsilon\rightarrow 0} \wp(\phi|\rho,\lambda,t) = \delta(t + \frac{\phi}{\omega} 
- \frac{\lambda}{\Omega}) 
\end{equation}
for all times, if this is the initial conditional probability.
For the non-deterministic case ~$\epsilon\ne0$, the probability distribution 
for the ~2+1-dimensional problem approaches a stationary distribution 
~$\wp_{as}(\phi|\rho,\lambda,t)$  
for large values of the time. So in this case, the ~$t \rightarrow \infty$ limit 
is taken before the ~$\epsilon \rightarrow 0$ limit. Then one finds that 
\begin{equation}
\lim_{\epsilon\rightarrow 0} \lim_{t\rightarrow \infty}\wp_(\phi|\rho,\lambda,t) = 
\lim_{\epsilon\rightarrow 0}\wp_{st}(\phi|\rho,\lambda) = \frac{1}{2\pi} . 
\end{equation}
Integrating eqn.(29) over ~$\phi$, and using eqn.(34) to calculate the 
conditional averages, we obtain for the marginal density 
~${\wp}(\rho,\lambda,t)$ 
\begin{eqnarray}
(\partial_t + \Omega\partial_\lambda){\wp}(\rho,\lambda,t)  
&=& - \frac{\partial}{\partial\rho}\Big[ \frac{\epsilon^{2b-1}Q}{2(\rho + 
{\hat r}_s\epsilon^{b+c-1})} 
+ \nu\hat F_0\epsilon^{b+3c-1} + \tilde\beta{\hat r}_s\epsilon^{b+3c-1} 
+ \tilde\beta\rho\epsilon^{2c} \nonumber\\ 
&-& \rho^3\Big( l - \beta'\frac{C_c}{\Delta^2}\frac{\mu}
{8 \cos^2\lambda}\Big)\epsilon^{2-2b} 
- 3\rho^2\Big( l - \beta'\frac{C_c}{\Delta^2}\frac{\mu}
{8 \cos^2\lambda}\Big){\hat r}_s\epsilon^{-b+c+1} \nonumber\\  
&-&3\rho\Big( l - \beta'\frac{C_c}{\Delta^2}\frac{\mu}
{8 \cos^2\lambda}\Big){{\hat r}_s}^2\epsilon^{2c} 
- \Big( l - \beta'\frac{C_c}{\Delta^2}\frac{\mu}
{8 \cos^2\lambda}\Big){{\hat r}_s}^3\epsilon^{b+3c-1} \Big] 
{\wp}(\rho,\lambda,t)\nonumber\\ 
&+& \frac{1}{2}\epsilon^{2b-1}\frac{\partial^2}{\partial\rho^2}Q{\wp}(\rho,\lambda,t) 
\end{eqnarray}
where the constants ~$\nu$ and  ~$\mu$ take the values 
\[ \nu = \left \{ \begin{array}{ll} 
1 & \mbox{~~for ~~ ~$\phi(t) = \Omega t + \alpha \pm 2n\pi$} \\
0 & \mbox{~~otherwise } \\
\end{array} 
\right .  \]  
\[ \mu = \left \{ \begin{array}{ll} 
2\cos^2\lambda & \mbox{~~when ~~ ~$\nu = 1$} \\
1 & \mbox{~~otherwise } \\
\end{array} 
\right .  \] 
and 
\begin{equation}
Q = Q_a + Q_b .
\end{equation}
We observe that the integration over ~$\phi$ immediately gives rise to an unexpected 
situation: {\em the driving term vanishes unless} ~$\phi = \lambda \pm 2n\pi$. Thus 
we are led to the important result that for an external periodic force to have any 
effect at all on the dynamics of a system with fluctuations near a Hopf bifurcation, 
the imaginary part (which is the fast variable) of the original two-dimensional 
Hopf oscillator (eqn.(3)), {\em must be phase-locked to the frequency of the external 
periodic force}. \\
For the system of the hair bundle, this translates to the condition that 
the ``effective active force'' of the hair bundle system or ``effective phase'' 
of the Hopf oscillator, ~$(\phi - \alpha)$ is in synchrony with the frequency 
of the periodic external stimulus: 
\begin{equation}
\phi(t) -\alpha = \Omega t \pm 2n\pi ~~~.
\end{equation}  

A little reflection shows that this situation stems from the existence of orbitally 
stable periodic solutions for the phase-locked condition in the absence of noise 
as in eqn.(23). \\

The drift and the diffusion terms in eqn.(35) should contribute equally 
to the probability density in the deterministic limit ~$(\epsilon\rightarrow 0)$ 
for all values of ~$c$ and thus should be of the same order in ~$\epsilon$. 
This necessitates that ~$b = 3/4$, ~from where one finds the value ~$c = 1/4$ 
~ separating the critical regime from the Gaussian one.  
Substituting these values into eqn.(35), we get
\begin{eqnarray}
\frac{1}{\epsilon^{1/2}}(\partial_t + \Omega\partial_\lambda){\wp}(\rho,\lambda,t) 
&=& -\partial_\rho\Big[ -2\rho\tilde\beta - 3{\tilde\beta}^{1/2}l^{1/2}\rho^2 - l\rho^3 
+ \frac{Q}{2\big(\rho + \frac{{\tilde\beta}^{1/2}}{l^{1/2}}\big)} 
+ \nu \hat F_0 \nonumber\\ 
&+& {\Big(\rho + \frac{{\tilde\beta}^{1/2}}{l^{1/2}}\Big)}^3\frac{\mu\beta'C_c}
{8\Delta^2\cos^2\lambda} \Big]{\wp}(\rho,\lambda,t)  + \frac{1}{2}Q\frac{\partial^2}
{\partial\rho^2}{\wp}(\rho,\lambda,t) .
\end{eqnarray}
The last term in the square bracket is the contribution from self-tuning. \\ 
The scaling of the time by ~$\epsilon^{-1/2}$ ~ on the left hand side shows that 
the dynamics of the critical, radial variable evolves at a different, slower 
time scale as 
compared with the phase variable, representing the reduction of the original 
system in the presence of noise to the center manifold ~$\frac{dC}{dt} = 0$ ~[36]. 
\\  

In general, even if a system is autonomous to begin with, a feedback control can 
turn it to a non-autonomous one. The normal form reduction for deterministic 
non-autonomous systems can be carried out as in [37] (see also [38,39])  
and in the limit of weak noise, its reduction to the normal form can be done as 
in [35,40].

Finally, having found the appropriate scaling for the variables, we now 
re-express eqn.(38) in terms of the original variables. We obtain  
\begin{eqnarray}
(\partial_t + \Omega\partial_\lambda){\cal P}(r,\lambda,t) &=& 
-\frac{\partial}{\partial r} 
\Big[ \beta r - \Big(l - \frac{\mu\beta'C_c}
{8\Delta^2\cos^2\lambda} \Big) r^3 + \nu F_0 
\Big]{\cal P}(r,\lambda,t)  \nonumber\\ 
&-& \frac{\partial}{\partial r}\Big( \frac{Q\epsilon}{2r} \Big){\cal P}(r,\lambda,t)   
+ \epsilon\frac{Q}{2}\frac{\partial^2}{\partial r^2}{\cal P}(r,\lambda,t) \nonumber\\
&=& ( L_0 + L_I ){\cal P}(r,\lambda,t) 
\end{eqnarray} 
where
\begin{eqnarray}
L_0 &=& -\frac{\partial}{\partial r} \Big[ \beta r - \Big(l - \frac{\mu\beta'C_c}
{8\Delta^2\cos^2\lambda} \Big) r^3 \Big] - 
\frac{\partial}{\partial r}\Big( \frac{Q\epsilon}{2r} \Big)  
+ \epsilon\frac{Q}{2}\frac{\partial^2}{\partial r^2} \nonumber\\
L_I &=& -\nu F_0 \frac{\partial}{\partial r} 
\end{eqnarray}
Using equations (29) to (35), the probability distribution for the full 
system is given by 
\begin{equation}
P(r,\phi,\lambda,t) = {\cal P}(\phi|r,\lambda) {\cal P}(r,\lambda,t) 
\end{equation}
Thus, as a first step, we have reduced the original 2+1-dimensional autonomous 
Hopf system (equation 29) to a 1+1-dimensional problem, {\em i.e.}, to a 
1-dimensional system in the slow variables in the extended phase space. 
The self-tuned system differs from the system without feedback in that the nonlinear 
part is modified in the drift term. \\ 
$L_0$ denotes the unperturbed Fokker-Planck operator. One would normally 
expect the periodic time dependence to be visible in the perturbation 
term $L_I$, ~but in the centre-manifold reduction procedure the averaging over 
the fast variable has eliminated this time-dependence through phase-locking. 
We observe another interesting phenomenon. 
The self-tuning term exhibits the periodic time-dependence in the Fokker-Planck 
equation which is absent only: ~ (i) in the presence of an external force 
and ~(ii) when there is no feedback in the system. This is seen by substituting 
the values of $\mu$ and $\nu$ from eqn.(36) into eqn.(39) and comes about 
because of the averaging done over the fast variable which is entrained with 
the external driving frequency and rotates with it. In the absence of external 
forcing, the system is oscillatory with intrinsic frequency $\omega$ and the 
$x(t)$ variable has a mode expansion with respect to this intrinsic oscillator 
frequency. In the unforced system with the kind of feedback we have considered 
(eqn.(4)), the control parameter has a similar mode expansion (because 
of eqns.(A-1) and (A-3)). As the fast variable for the unforced system is not 
phase-locked with any frequency, averaging over the fast variable does not 
eliminate the time periodicity in the Fokker-Planck equation. 
{\em Thus, even in the absence 
of an external force, the effect of a feedback as in equations (4) and (A-3) 
is felt as a periodic modulation in the full Fokker-Planck equation}.\\

We assume the following physical boundary condition for $r$: 
~${\cal P}(r\rightarrow \infty,\lambda,t) = 0$. 
Since ~$\lambda$ is periodic, ~~${\cal P}_t(r,\lambda) = {\cal P}_t(r,\lambda + 2\pi)$, 
we can perform a Fourier expansion of ~${\cal P}(r,\lambda,t)$ in ~$\lambda$ to 
find the general asymptotic solution of eqn.(39) 
\begin{equation}
{\cal P}_{as}(r,\lambda,t) = \sum_{n=-\infty}^{+\infty}W_n(r,\lambda)e^{in\Omega t} = 
\sum_{n=-\infty}^{+\infty} w_n {\cal P}_{st}(r,\lambda)
e^{-in(\lambda -\Omega t)}
\end{equation} 
where  ~${\cal P}_{st}$  is the solution of the stationary problem for uniformly 
distributed phases ~$\alpha$ and the Fourier coefficients obey
\begin{equation}
\Big[L_0 + L_I - \Omega \frac{\partial}{\partial \lambda} 
- in\Omega \Big]W_n(r,\lambda) = 0 ,
\end{equation} 
$w_n$ ~denoting the weight of the initial probability ~${\cal P}(r,\lambda,t=0)$ 
on the eigenfunction ~$e^{-in\lambda}$ with eigenvalue ~$in\Omega$.

\subsection*{4. Adiabatic limit} 

We solve for the probability distribution for the simplest case, for very 
small values of the frequency. 
In this limit, ~$\lambda$  varies very slowly with time so that we can consider the 
simplified problem  
\begin{eqnarray}
\partial_t{\cal P}_{ad}(r,\lambda,t) &=& -\frac{\partial}{\partial r} \Big[ \beta r 
- \Big( l - \frac{\mu\beta'C_c}
{8\Delta^2\cos^2\lambda} \Big)r^3 + \nu F_0 \Big]{\cal P}_{ad}(r,\lambda,t) \nonumber\\  
&-& \frac{Q\epsilon}{2}\frac{\partial}{\partial r}
\Big( \frac{1}{r} \Big){\cal P}_{ad}(r,\lambda,t)   
+ \frac{Q\epsilon}{2}\frac{\partial^2}{\partial r^2}{\cal P}_{ad}(r,\lambda,t) .
\end{eqnarray}
\noindent For this case, we find that the solution of the stationary problem for 
this situation is  
\begin{equation}
\lim_{t\rightarrow \infty}{\cal P}_{ad}(r,\lambda,t) = {\cal N}_e r 
\exp{\Big\{ -\frac{2}{Q\epsilon}
\Big[ (l - \frac{\mu\beta'C_c}{8\Delta^2\cos^2\lambda})\frac{r^4}{4} 
- \beta\frac{r^2}{2} - \nu F_0r \Big] \Big\}}
\end{equation}

\noindent where the normalization constant ~${\cal N}_e$ is given by  
\begin{equation}
{\cal N}_e = {\cal N}_0{\Big \{ \sum_{n=0}^\infty  
\frac{\Gamma(\frac{n}{2} + \frac{3}{2})}{n!}{\Big(\frac{2\nu F_0}
{{(Q\epsilon)}^{\frac{3}{4}}{\big(l - \frac{\mu\beta'C_c}{8\Delta^2\cos^2\lambda}\big)}
^{\frac{1}{4}}}\Big)}^n  
{\cal D}_{-\frac{n}{2} -\frac{3}{2}}
\Big(\frac{-\beta}
{{[Q\epsilon(l - \frac{\mu\beta'C_c}{8\Delta^2\cos^2\lambda})]}^{\frac{1}{2}}}\Big) 
\Big \}}^{-1} ,
\end{equation}
where 
\begin{equation}
{\cal N}_0 =  4\pi{\Big(\frac{(l - \frac{\mu\beta'C_c}{8\Delta^2\cos^2\lambda})}
{Q\epsilon} \Big)}^{\frac{3}{4}}
\exp{\Big[ - \frac{\beta^2}{4Q\epsilon
(l - \frac{\mu\beta'C_c}{8\Delta^2\cos^2\lambda})}\Big] }
\end{equation}
and ~${\cal D}_{-n}$ are parabolic cylinder functions. \\ 

When the fast variable is phase-locked to the external periodic force we 
obtain the time-independent stationary probability density for the 
forced system to be 
\begin{equation}
\lim_{t\rightarrow \infty}{\cal P}_{ad}(r,\lambda,t) = {\cal N}_F r 
\exp{\Big\{ -\frac{2}{Q\epsilon}
\Big[ (l - \frac{\beta'C_c}{4\Delta^2})\frac{r^4}{4} 
- \beta\frac{r^2}{2} - F_0r \Big] \Big\}}
\end{equation} 
where 
$${\cal N}_F =  4\pi{\Big(\frac{(l - \frac{\beta'C_c}{4\Delta^2})}
{Q\epsilon} \Big)}^{\frac{3}{4}}
\exp{\Big[ - \frac{\beta^2}{4Q\epsilon
(l - \frac{\beta'C_c}{4\Delta^2})}\Big] }
{\Big \{ \sum_{n=0}^\infty  
\frac{\Gamma(\frac{n}{2} + \frac{3}{2})}{n!}{\Big(\frac{2 F_0}
{{(Q\epsilon)}^{\frac{3}{4}}{\big(l - \frac{\beta'C_c}{4\Delta^2}\big)}
^{\frac{1}{4}}}\Big)}^n  
{\cal D}_{-\frac{n}{2} -\frac{3}{2}}
\Big(\frac{-\beta}
{{[Q\epsilon(l - \frac{\beta'C_c}{4\Delta^2})]}^{\frac{1}{2}}}\Big) 
\Big \}}^{-1}$$ . 

Thus even in the presence of external driving, the probability density 
evolving on the time scale of the slower critical variable is 
stationary because the fast variable over which averaging has been 
performed is phase locked with the driving frequency and rotates with it. 

Fig.1 shows the effect of self-tuning on a plot of the probability 
distribution as a function of the radial variable. The presence of the 
self-tuning term increases the width of the distribution, increases the 
height of its peak and moves it to the right.  
This is enhanced much more in a forced system since the driving term 
increases the overall heights of the peaks and shifts these more towards 
the right (than for the unforced system). This is seen in Fig.2 which is the  
solution of eqn.(44) for the forced system in the long time limit. The 
sensitivity of the distribution to the angle is much larger at its decreasing 
side after the peak, the probability distribution not being symmetric.  \\  
The trajectories in the x-y plane for the original system (eqn.11) (written in 
terms of polar coordinates in eqn.(17)) are shown in a representative 
simulation (Fig.3). 
The closed orbit of 
unforced deterministic Hopf oscillator is shown in the solid line while the 
dashed orbit is for the deterministic Hopf oscillator perturbed by an 
external driving force. The dotted curve shows how the path gets completely 
changed in the presence of a very small additive noise. The effect of the 
feedback we have considered is seen in the figure (dash-dotted orbit) which 
``controls'' the deviation of the path of the noisy system, bringing it 
towards the deterministic trajectory. \\    
 
In the absence of an external driving force (when $\nu = 0$) or when the 
condition of phase-locking to the external force eqn.(37) is not satisfied, 
we observe a new feature --- {\em the long-time limit of the probability 
distribution for the self-tuned 
unforced system is not stationary}: 
\begin{equation}
\lim_{t\rightarrow \infty}{\cal P}_{ad}(r,t) = {\cal N} r 
\exp{\Big\{ -\frac{2}{Q\epsilon}
\Big[ (l - \frac{\beta'C_c}{8\Delta^2\cos^2(\omega t + \alpha)})\frac{r^4}{4} 
- \beta\frac{r^2}{2} \Big] \Big\}} .
\end{equation} 
In the above expression,
\begin{equation}
{\cal N} = \frac{2}{\pi^{\frac{1}{2}}}{\Big(
\frac{(l - \frac{\beta'C_c}{8\Delta^2\cos^2(\omega t + \alpha)})}
{Q\epsilon} \Big)}^{\frac{3}{4}}
\exp{\Big[ - \frac{\beta^2}{4Q\epsilon
(l - \frac{\beta'C_c}{8\Delta^2\cos^2(\omega t + \alpha)})}\Big] }
{\Big({\cal D}_{-\frac{3}{2}}
\Big(\frac{-\beta}
{{[Q\epsilon(l - \frac{\beta'C_c}{8\Delta^2\cos^2(\omega t + \alpha)})]}^
{\frac{1}{2}}}\Big)\Big)}^{-1} .
\end{equation}

Thus, a stationary probability distribution in the strict sense of 
time-independence does not exist even in the absence of all external forces 
for ~$\nu = 0$, when there is a feedback in the system in which the control 
parameter is generated by the system dynamics such as in equations 
(4), (A-1) and (A-3). 
In the adiabatic limit, the large time distribution is periodically 
modulated by the self-tuning term, always with ~$l > \frac{\beta'C_c}
{8\Delta^2\cos^2(\omega t + \alpha)}$ , ~~and exhibits peaks close to 
\begin{equation}
t \approx < \frac{1}{\omega}\Big( \cos^{-1}{\big(\frac{\beta'C_c}
{8\Delta^2l}\big)}^{1/2} 
- \alpha \Big) \pm \frac{2\pi}{\omega}n  ~~~, ~~~~n = 0, 1, 2, 3, \dots 
\end{equation}   
These features are absent when there is no feedback control on the system. \\ 

The time-periodicity of the probability density for the self-tuned 
system in the absence of external 
force can be understood if we bear in mind that we are studying the 
behaviour of a system which can maintain self-sustained oscillations 
(even in the absence of external driving), at the onset of a limit cycle. 
The feedback term in the self-tuned system makes the system explicitly 
non-autonomous even in the absence of external driving and the noise 
averages of quantities, such as correlation functions are calculated in 
this case as for a quasistationary process, just as for the case of a 
driven system. But the unforced self-tuned system differs from the driven 
system in that the fast variable is not entrained with a particular frequency, 
in this case with the intrinsic frequency of the limit cycle. 
Then an averaging performed over the fast 
variable does not eliminate the temporal periodicity, and this is 
reflected correspondingly in the probability density. \\ 
It is known since the work of [42] that dynamical systems (both autonomous 
and non-autonomous) showing such periodic behaviour close to instabilities 
show an enhanced output in the power spectrum from the transient response 
before the bifurcation due to an increase in the relaxation time 
when the instability is approached, a phenomenon termed coherence resonance. 
We elaborate a little more on this for our system, towards the end of 
this section.\\ 

In the adiabatic limit, the mean value ~$\langle x(t)\rangle_{ad}$  of ~$x(t)$ 
for the forced system is given by 
\begin{eqnarray}
\langle x(t, \alpha)\rangle_{ad} &=& \cos(\Omega t +\alpha \pm 2k\pi)
\frac{{(Q\epsilon)}^{\frac{1}{4}}}
{{(l - \frac{\beta'C_c}{4\Delta^2})}^{\frac{1}{2}}}\nonumber\\
&\times& \frac{ \sum_{n=0}^\infty  
\frac{(n+1)(\frac{n}{2} + 1)}{\Gamma(\frac{n}{2} + \frac{3}{2})}{\Big(\frac{F_0}
{{(Q\epsilon)}^{\frac{3}{4}}{\big(l - \frac{\beta'C_c}{4\Delta^2}\big)}
^{\frac{1}{4}}}\Big)}^n 
{\cal D}_{-\frac{n}{2} -2}
\Big(\frac{-\beta}
{{[Q\epsilon(l - \frac{\beta'C_c}{4\Delta^2})]}^{\frac{1}{2}}}\Big)}
{ \sum_{n=0}^\infty  
\frac{(n+1)}{\Gamma(\frac{n}{2} + 1)}{\Big(\frac{F_0}
{{(Q\epsilon)}^{\frac{3}{4}}{\big(l - \frac{\beta'C_c}{4\Delta^2}\big)}
^{\frac{1}{4}}}\Big)}^n 
{\cal D}_{-\frac{n}{2} -\frac{3}{2}}
\Big(\frac{-\beta}
{{[Q\epsilon(l - \frac{\beta'C_c}{4\Delta^2})]}^{\frac{1}{2}}}\Big)}
\end{eqnarray}
Details of the derivation of eqn.(52) are given in Appendix B. 
In the weak noise limit the asymptotic expansions of the parabolic cylinder functions 
may be used [41]. A few terms of the expansion are displayed below 
\begin{eqnarray}
\langle x(t, \alpha)\rangle_{ad} &=& \frac{{\bar r}_s^{\frac{1}{2}}}{\beta^{\frac{3}{4}}}
\cos(\Omega t +\alpha \pm 2k\pi)\nonumber\\
&\times& \frac{\Big \{ \frac{1}{{(Q\epsilon)}^{\frac{1}{2}}}
\Big[ (\beta^{\frac{1}{2}}{\bar r}_s 
+ \frac{3F_0}{4\beta^{\frac{1}{2}}}) - \frac{F_0^2{\bar r}_s}{2\beta^{\frac{1}{2}}
Q\epsilon} 
-  \frac{F_0\beta^{\frac{1}{2}}{\bar r}_s^3}{2{(Q\epsilon)}^2} + \dots \Big] 
+ \frac{2F_0}{Q\epsilon}\Big( {\bar r}_s + \frac{F_0^2{\bar r}_s}{4\beta Q\epsilon} + 
\frac{2F_0^2}{3{(Q\epsilon)}^2} + \dots \Big)\Big \}}
{ \Big \{ \frac{1}{{(Q\epsilon)}^{\frac{1}{2}}}\Big[ (\beta^{\frac{1}{2}}{\bar r}_s 
+ \frac{3F_0^2{\bar r}_s}{4\beta^{\frac{1}{2}}Q\epsilon}) 
+ \frac{F_0^2{\bar r}_s\beta^{\frac{1}{2}}}{{(Q\epsilon)}^2}(2{\bar r}_s^2 + 
\frac{5F_0^2}{4\beta^2}) +\dots \Big]  
+ \frac{2F_0}{Q\epsilon}{\bar r}_s\Big( 1 -\frac{2F_0^2}{3\beta Q\epsilon}  
- \frac{2F_0^2{\bar r}_s^2}{3{(Q\epsilon)}^2} +\dots \Big)\Big\}}\nonumber\\
\end{eqnarray}
where ~${\bar r}_s$ is defined in eqn.(23). \\ 
We find the response of the system in the weak noise limit following [24,25] by 
expanding ~$\langle x(t, \alpha)\rangle_{ad}$ in a Fourier series  
\begin{equation}
\langle x(t, \alpha)\rangle_{ad} = \sum_{-\infty}^{\infty}M_n e^{in(\Omega t + \alpha)} 
\approx 2|M_1|\cos(\Omega t +\alpha)
\end{equation}
A plot of the response versus the noise strength in the weak noise limit 
shows a non-monotonic behaviour, with very steep peaks at certain values 
of the noise strength, reminiscent of stochastic resonance (Fig.4). \\
It is also possible to compute the resulting spectrum in the asymptotic limit 
following [25], from the phase-averaged asymptotic correlation function 
~${\bar K}_{as}(t,t')$ 
\begin{equation}
{\bar K}_{as}(t,t') = 2\sum_{n=1}^{\infty}{|M_n|}^2\cos n\Omega t 
\end{equation}
The asymptotic spectral density which contains the delta spikes is defined from 
\begin{equation}
{\bar S}_{as}(\psi) = \int_{-\infty}^{\infty}d\tau {\bar K}_{as}(\tau)e^{-i\psi\tau} 
= 2\pi\sum_{n=1}^{\infty}{|M_n|}^2 \delta(\psi - n\Omega) . 
\end{equation}
Also as in [25], the spectral power amplification ~$\eta$ can be calculated by 
dividing the power ~$P_n$ in the $n$th frequency component by the total input power  
~$P_{0}$ in the modulation  
\begin{equation}
\eta = \frac{P_1}{P_0} =  \frac{4\pi{|M_1|^2}}{\pi F_0^2} = 4{(\frac{|M_1|}{F_0})}^2 .
\end{equation}

We observe that ~$\eta$ depends upon the amplitude for each frequency (Fig.5). Because 
of phase-locking, the frequency dependence is not apparent in our formalism 
employing centre-manifold reduction. As the amplitude of the forcing increases,  
the amplification of the response of the system moves towards increasing values 
of the noise strength, diminishing rapidly in magnitude (Fig.5). Thus, {\em a noisy 
Hopf oscillator amplifies better weak signals}. This is reminiscent of the response of 
the ear --- --- the ear is more responsive to weak signals, cochlear amplification being 
most pronounced at the auditory threshold and falling steeply with increasing 
stimulus intensity [5,10].  \\
 
It is very important to note, however, that the underlying phenomenon in a noisy 
Hopf oscillator with external periodic forcing is completely different from that 
in stochastic resonance, and rather unrelated to it. In a Hopf oscillator, 
even though the periodic forcing is imposed from outside, in the 1:1 resonance 
we have considered here, the oscillator has the same frequency of oscillation 
as the external periodic force. 
More important, in the presence of noise, there is phase-locking of the fast 
variable of the oscillator with the external frequency. Thus, there are no 
oscillations in the sense of stochastic resonance. However, in this case also, 
the sharply enhanced response appears to be the consequence of the interaction 
of the two kinds of perturbations with the stable periodic orbits of the 
unperturbed system, the noise  always trying to kick the system out of the 
orbit while the applied phase-locked force kicks it back again towards 
the orbit. \\

In the presence of noise, the bifurcation point is disturbed  
(see [1] \& references therein). In order to tune 
the system close to the critical point, the distance of the control parameter from 
its critical value ~$C_c$ must be minimal  
\begin{equation}
\delta C = C - C_c \approx \frac{C_c}{\Delta^2}{|x_1|}^2 \rightarrow 0 .
\end{equation}
For the system with noise, the quantity of interest is the noise averaged distance 
~$\langle\delta C\rangle$ from the critical point. \\  
In the presence of an external stimulus, we find on using eqn.(16) 
\begin{eqnarray}
\langle \delta C \rangle_e &\approx& \frac{C_c}{4\Delta^2 \cos^2(\Omega t + \alpha)}
\frac{{\cal N}_F}{2\pi}
\int_0^{2\pi} d\phi \int_0^\infty dr r^4 \cos^2\phi \delta(\phi-\Omega t 
-\alpha \pm 2n\pi)
\exp{\Big\{ -\frac{2}{Q\epsilon}
\Big[ (l - \frac{\beta'C_c}{4\Delta^2})\frac{r^4}{4} 
- \beta\frac{r^2}{2} - F_0r \Big] \Big\}}\nonumber\\ 
&\approx& \frac{C_c}{4\Delta^2}
\frac{{(Q\epsilon)}^{\frac{1}{4}}}
{2{(l - \frac{\beta'C_c}{4\Delta^2})}^{\frac{1}{2}}} 
\frac{ \sum_{n=0}^\infty  
\frac{(n+1)(n+3)}{\Gamma(\frac{n}{2} + 1)}{\Big(\frac{F_0}
{{(Q\epsilon)}^{\frac{3}{4}}{\big(l - \frac{\beta'C_c}{4\Delta^2}\big)}
^{\frac{1}{4}}}\Big)}^n 
{\cal D}_{-\frac{n}{2} -\frac{5}{2}}
\Big(\frac{-\beta}
{{[Q\epsilon(l - \frac{\beta'C_c}{4\Delta^2})]}^{\frac{1}{2}}}\Big)}
{ \sum_{n=0}^\infty  
\frac{(n+1)}{\Gamma(\frac{n}{2} + 1)}{\Big(\frac{\nu F_0}
{{(Q\epsilon)}^{\frac{3}{4}}{\big(l - \frac{\beta'C_c}{4\Delta^2}\big)}
^{\frac{1}{4}}}\Big)}^n 
{\cal D}_{-\frac{n}{2} -\frac{3}{2}}
\Big(\frac{-\beta}
{{[Q\epsilon(l - \frac{\beta'C_c}{4\Delta^2})]}^{\frac{1}{2}}}\Big)}\nonumber\\
&\approx& \frac{C_c \beta}{4\Delta^2{(l - \frac{\beta'C_c}{4\Delta^2})}^{\frac{3}{2}}}
\frac{[1 - \frac{4F_0}{{(\beta Q\epsilon)}^{\frac{1}{2}}} 
+ 2\frac{F_0^2}{\beta Q\epsilon} 
- \frac{4}{3}{(\frac{F_0}{{(\beta Q\epsilon)}^{\frac{1}{2}}})}^3 +\dots]}
{[1 + \frac{2F_0}{{(\beta Q\epsilon)}^{\frac{1}{2}}} 
+ 2\frac{F_0^2}{\beta Q\epsilon} 
+ \frac{4}{3}{(\frac{F_0}{{(\beta Q\epsilon)}^{\frac{1}{2}}})}^3 +\dots]} .
\end{eqnarray}
In this case, the feedback reduces the real part of the cubic term 
of the original Hopf system by a constant amount. Also ~$\langle\delta C\rangle_e$ 
increases with the noise, first rapidly, and then gradually plateaus off for larger 
values of the noise (Fig.6). \\ 

For the unforced self-tuned system, we obtain 
\begin{eqnarray}
\langle \delta C \rangle &\approx& \frac{C_c}{4\Delta^2 \cos^2(\omega t + \alpha)}
\frac{\cal N}{2\pi}
\int_0^{2\pi} d\phi \int_0^\infty dr r^4 \cos^2\phi 
\exp{\Big\{ -\frac{2}{Q\epsilon}
\Big[ (l - \frac{\beta'C_c}{8\Delta^2\cos^2(\omega t + \alpha)})\frac{r^4}{4} 
- \beta\frac{r^2}{2} \Big] \Big\}} \nonumber\\
&\approx& \frac{3C_c}{64\pi\Delta^2\cos^2(\omega t +\alpha)}{\Big(\frac{Q\epsilon}
{l - \frac{\beta'C_c}{8\Delta^2\cos^2(\omega t + \alpha)}}\Big)}^{\frac{1}{2}}
\frac{{\cal D}_{-\frac{5}{2}}\Big(\frac{-\beta}
{{[Q\epsilon(l - \frac{\beta'C_c}{8\Delta^2\cos^2(\omega t + \alpha)})]}^
{\frac{1}{2}}}\Big)}
{{\cal D}_{-\frac{3}{2}}\Big(\frac{-\beta}
{{[Q\epsilon(l - \frac{\beta'C_c}{8\Delta^2\cos^2(\omega t + \alpha)})]}^
{\frac{1}{2}}}\Big)}\nonumber\\
&\approx& \frac{C_c}{32\pi\Delta^2}\frac{{\bar r}_0^2}{\cos^2(\omega t +\alpha)} 
\Big( 1 + \frac{Q\epsilon}{2\beta}\frac{1}{{\bar r}_0^2} 
+ \frac{15}{16}{(\frac{Q\epsilon}{2\beta})}^2
\frac{1}{{\bar r}_0^4} + \dots \Big) 
\end{eqnarray}
where ~~~
\begin{equation}
{\bar r}_0 = {\Big(\frac{\beta}
{l - \frac{\beta'C_c}{8\Delta^2\cos^2(\omega t + \alpha)}}\Big)}^{\frac{1}{2}} 
~~~~~~~~~~~, ~~~~~~
\end{equation}
and asymptotic expansions of the parabolic cylinder functions have been used 
in the weak noise limit as before. In this limit, the contributions from higher 
order noise terms progressively diminish and can be neglected. We observe that the 
contribution from the feedback equation for a fixed noise strength modulates 
~$\delta C$ periodically (Fig.7), spiking at specific frequencies. Thus, 
when one tries to tune the system close to the critical point for a 
given noise strength, at certain frequencies of the oscillator, it suddenly 
departs, far from the bifurcation. This behaviour essentially arises because 
of the dependence of the control parameter on the system dynamics. The large 
departures at certain frequencies reflect the role of the fluctuations in 
perturbing the system away from the stable limit cycle.\\ 

\noindent The sign of ~$\langle \delta C \rangle$ would determine whether the 
fluctuations have delayed or advanced the bifurcation. It is known in general 
that fluctuations smear and advance the bifurcation in an uncoupled system. 
For a generic resonantly forced Hopf oscillator, we see from eqn.(59) that 
~$\langle \delta C \rangle$ would always be positive. In an unforced system 
subject to the feedback in eqn.(4) however, we see from eqn.(60) that the sign 
of ~$\langle \delta C \rangle$ would depend upon whether ~$l$ is greater 
than or less than ~$\frac{\beta'C_c}{8\Delta^2\cos^2(\omega t + \alpha)}$. \\
To tune any dynamical system to operate close to the Hopf bifurcation in 
the presence of fluctuations, one would need to adjust the values of the 
parameters at the operating point so that 
~$\langle \delta C \rangle \rightarrow 0$. For a nonlinear system to act 
as an efficient detector of signals in the presence of noise, it s necessary 
to choose parameters appropriately so that the system neither gets subject 
to large oscillations, nor is it far into the quiescent regime, but rather 
operates as close to the critical point as possible in order to achieve 
the maximal value of the spectral power amplification for the signal to 
be amplified effectively. \\
If indeed the mechanotransducers in the inner ear, the hair cells, are 
Hopf oscillators {\em in vivo}, then, it appears that Nature has designed 
the living system in such a way that despite numerous defects and anomalies 
which occur in the organism, the actual biological parameters --- the 
stiffness of the hair bundle, the length of the stereocilia, the calcium 
concentration entering them, etc., are all so accurately regulated to 
operate close to the oscillatory instability, as to achieve optimal 
efficiency and detect even the faintest whisper. \\ 

\noindent Finally, we would like to mention that the spectrum of the unforced 
self-tuned system shows very interesting behaviour. Because of the nature 
of the feedback in equations (4),(A-1) and (A-3), the self-tuned system is 
periodic in time and would therefore, in the presence of noise, be expected 
to exhibit the characteristic precursor of the Hopf instability [42]. 
Noisy precursors of instabilities in nonlinear systems were first studied 
and classified systematically in [42] and subsequently noise enhancement 
of precursors has been demonstrated in [43]. 
   
Using the phase-averaged asymptotic correlation function 
~${\bar K}_{as}(t,t') = \langle\langle x(t) x(t')\rangle\rangle_\alpha$ ~ and 
after some simplifications, the asymptotic spectral density has the form 
(to lowest order approximation) 
\begin{eqnarray}
{\bar S}_{as}(\psi) &=& \int_{-\infty}^{\infty}d\tau {\bar K}_{as}(\tau)
e^{-i\psi\tau} 
\nonumber\\
&\approx&  \int_{-\infty}^{\infty}d\tau \frac{1}{8\pi}
\Big( \frac{Q\epsilon}{2\beta} + h(t) \Big) e^{-i(\psi - n\omega)\tau} .
\end{eqnarray}
Here, ~$\tau = t-t'$ ~and the time periodicity of ~${\bar K}_{as}(\tau)$ 
arising from the feedback (the self-tuning terms) is used to expand it in 
Fourier series. The time-dependent periodic part has been represented 
by  ~$h(t) = h(t+T)$. Approximations such as in eqn.(60) are made in the 
evaluation of the ratio of the parabolic cylinder functions 
~$\frac{{\cal D}_{-\frac{5}{2}}(-u)}{{\cal D}_{-\frac{3}{2}}(-u)}$. ~
We can see from here that the spectrum consists of delta peaks at frequencies 
~$(\psi-n\omega), ~n=0,1,2,\dots $ ~superimposed on the spectrum of a bounded 
time-periodic function. The latter arises because of the oscillatory nature 
of the feedback. \\
The delta peaks can be identified with contributions from the nonzero Floquet 
exponents, with the real part of the Floquet exponent (~$Re(A(C_c))=\beta$) 
determining the size and shape of the peaks and the position of the peaks being 
determined by the imaginary part of the exponents (~$Im(A(C_c))=\omega$). \\ 
The sign of the real part of the Floquet exponents determines the stability 
of the orbit in phase space, a stable orbit being characterized by negative 
values for all the Floquet multipliers [42], a change in stability being 
indicated by a change in their sign, the stability of the system therefore 
being dominated by the properties of the Floquet exponent with the smallest 
negative real part. 
This is captured by the Lamrey diagrams for the successive iterates of the 
orbit in phase space, in its corresponding Poincare return map [44]. 
A statistical method based on the properties of the behaviour of the 
return map was developed in a recent work [45] to detect the onset of 
bifurcations and their precursors.\\ 

\noindent Our results and observations above allow us to go a step further 
and speculate on the implications, in the biological context of hearing,
of feedback for an unforced Hopf oscillator subject to fluctuations. \\ 
We suggest that in the context of the Hopf oscillator hypothesis for the 
sensory hair cells, the self-tuning mechanism for a noisy Hopf oscillator 
could be responsible, to a large extent, for the sharp peaks observed 
in the spectra of 
otoacoustic emissions emitted spontaneously from the ears of various 
vertebrates [8]. This could also be the reason for the corresponding 
phenomenon of autonomous vibrations observed in Johnston's organ [9], the 
hearing organ of insects where the mechanotransducers are sensory neurons, 
rather than hair cells.  \\ 
The effect of the external noise on the self-tuning at the onset of an 
instability is thus visible as 
the sharp peaks in the spectrum of the otoacoustic emissions. \\ 
Of course, in order to determine the frequencies at which the spectrum 
exhibits the delta peaks for a given hair bundle (or for the sensory neurons 
in insects), one would need to know all the parameters describing its 
mechanical properties -- such as the length of the stereocilia, 
their stiffness, etc.. However, the range of the otoacoustic spectrum would 
be determined by various other factors such as the number of the 
mechanotransducers (Hopf oscillators), their characteristic frequencies, their 
arrangement on the basilar membrane, the length of the basilar membrane, the 
influence of their attachments if any (such as the tectorial membrane in many 
organisms), and also, possibly, the influence of the efferent neurons on them. 
The scope of our study, however, does not extend to working out the 
biological details of such a system.  

\subsection*{5. Conclusion} 

We have discussed how a Hopf oscillator in the presence of additive noise displays 
an amplified response in the presence of an external stimulus displaying some kind 
of a stochastic filtering effect. The effect is more pronounced for weaker signals 
and diminishes as the magnitude of the stimulus increases. \\ 
We have considered the case of a feedback which is generated by the system 
dynamics and have eliminated the fast oscillations of the system by means 
of centre manifold reduction, enabling us to solve the associated Fokker Planck 
equation in the adiabatic limit.\\
The fast variable of the Hopf oscillator gets phase-locked to the external frequency 
in the presence of periodic driving. \\ 
In the presence of an external periodic force, the distance from the bifurcation 
increases with increasing noise strength, first very steeply and then plateaus, 
showing a very gradual but definite increase with the noise. In the absence of 
external forces it shows a dependence on the frequency of the limit cycle for 
a fixed noise strength, exhibiting sharp peaks at specific frequencies.\\

\noindent The estimate found in our work for the distance from the bifurcation 
provides a useful measure for building, for instance, a detector for detecting 
weak signals in a noisy environment. The parameters of the actual specific 
dynamical system must be so adjusted as to have a minimal, if not an almost 
vanishing value for ~$\langle \delta C \rangle$  so that the system operates 
with maximal efficiency, and with the largest feasible value of the spectral 
power amplification so that even the faintest signals can be captured. \\ 
Using the actual values of biological parameters from biophysical models 
of the hair cell in our estimate would be a useful exercise and test to 
check the applicability of our theoretical treatment of self-tuning a 
Hopf oscillator, to actual biology. \\   
 
\noindent The phenomenon of stochastic resonance has been proposed as a 
possible underlying mechanism for amplification of weak signals by the ear 
[21,16]. Our analysis shows that if the underlying force producing elements 
in the inner ear are indeed Hopf oscillators as has been proposed by several 
authors [10-12], then the reason 
for the amplified response is not because of stochastic resonance, and the 
totally unrelated stochastic ``filtering'' effect which we have demonstrated 
here might be the underlying cause. \\ 
The important feature which distinguishes these two phenomena is that of 
entrainment of the phase of the noisy Hopf oscillator with the stimulus 
frequency, which does not happen for stochastic resonance. \\
While for the forced Hopf oscillator the peaks in the spectral power 
amplification diminish in magnitude with increasing stimulus amplitude, 
they do so moving towards increasing values of the noise strength. 
In contrast, in the case of stochastic resonance for a bistable potential, this 
shift is towards decreasing values of the noise strength [25].  
These are results which are also amenable to experimental analysis. 
A key point of course is that stochastic resonance does not occur in the 
absence of an external stimulus. \\
We found at the end of Section 4  that an unforced Hopf oscillator 
with feedback, does, however show the expected coherence resonance. \\ 

\noindent We suggest that within the framework of the Hopf oscillator hypothesis 
in hearing research, the sharp peaks observed in the spectrum of spontaneous 
otoacoustic emissions of various organisms could be a signature of the 
noisy precursor to the Hopf instability for the unforced self-tuned oscillator. \\  
Of course, several alternative views and approaches have been extant  
in hearing research [46], and various theories have been proposed for the 
occurrence of otoacoustic emissions (both spontaneous and evoked) --- see, for 
instance, ref.[47] and references therein. The description of the individual hair 
cells arranged on the cochlea as Hopf oscillators responding to stimulus 
frequencies resonant with their individual characteristic frequencies is 
fairly recent and interesting, and a description of the hair cell in a 
generic way aims to broadly capture the essential dynamics common 
to that in all organisms --- mammals as well as non-mammals. 
In this framework, our analysis could give useful insight into the effect 
of fluctuations on regulatory feedback mechanisms at the cellular level.\\  

To our knowledge, our work provides the first theoretical analysis  
on the consequences of feedback in a Hopf oscillator subject to noise. \\ 

A more physically appealing system to consider would be one in which the 
control parameter fluctuates --- the resulting multiplicative noise could 
give much richer effects. Studies in this direction are under progress and 
will be reported elsewhere.\\  

On the other hand, we would like to point out that when the system under 
study is a periodically forced Hopf oscillator, care must be taken to ensure 
that the correct equation (3) for the normal form [28-31] is employed 
to model it. 
This is especially essential when the external forcing contains more 
than one frequency. Thus it would be instructive to repeat the analysis 
in [14] using the well-established normal form equation (3) for a 
periodically forced system near a Hopf bifurcation [28-31]. \\

\subsection*{Appendix A: Normal mode equation for a Hopf oscillator }
\renewcommand{\theequation}{A-\arabic{equation}} 
\setcounter{equation}{0}
We state here the essential outline of the steps followed to derive eqn.(5).
For a periodic force ~$F_{ext}(t) = F_{ext}(t + \frac{2\pi}{\Omega})$, ~$x$ can be 
expanded in terms of its Fourier modes: ~$x(t) = \sum_n x_n e^{in\Omega t}$. Since 
we consider a feedback such as in eqn.(4) which depends upon ~$x(t)$, we can expand 
$C(t)$ in Fourier modes  
\begin{equation}
C(t) = C_0(t) + \sum_{n \ne 0} C_n e^{i n\Omega t}
\end{equation} 
where: ~$C_{-n} = C_n^*$. 

As in [11], a solution of eqn.(4) can be found by inverting $F$ and expressing 
it as a polynomial in the variable ~$x(t)$ and ~$C(t)$. In Fourier components, 
we write this as   
\begin{eqnarray}
F_k &=& {\cal F}_{kl}^{(1)}x_l + {\cal F}_{klm}^{(2)}x_lx_m + 
{\cal F}_{klmn}^{(3)}x_lx_mx_n + 
G_{kl}^{(1)}C_l + G_{klm}^{(2)}C_lC_m + G_{klmn}^{(3)}C_lC_mC_n \nonumber \\ 
&+& H_{klm}^{(2)}x_lC_m + H_{klmn}^{(3)}x_lx_mC_n +  M_{klmn}^{(3)}x_lC_mC_n 
+ \dots 
\end{eqnarray} 
The indices ~$k_i$ ~in the expansion coefficients  ~${\cal F}_{k, k_1,\dots, 
k_n}^{(n)}$, ~$G_{k, k_1,\dots, k_n}^{(n)}$, etc., are constrained by: 
~$k = k_1 + \dots +k_n$.  
All expansion coefficients are symmetric with respect to permutations 
of ~$k_i$.\\  

The dependence of the control parameter on the $x$ variable can 
then be expressed in a general way, as a power series in $x$   
\begin{equation} 
C_k = \zeta_{kl}^{(1)}x_l + \zeta_{klm}^{(2)}x_lx_m + 
\zeta_{klmn}^{(3)}x_lx_mx_n + \dots ~~, ~~~~k \ne 0 
\end{equation}  
where the expansion coefficients ~ $\zeta_{k, k_1,\dots, k_n}^{(n)}$ are 
functions of the frequency ~$\omega$ of the limit cycle of the oscillator, 
and symmetric with respect to permutations of the indices ~$k_i$  which 
satisfy: 
~~ $k = k_1 + \dots +k_n$.  \\ 
Substituting eqn.(A-3) into eqn.(A-2) leads to a polynomial equation in $x_0$, $x_k, 
(k\ne0)$, and $C_0$. ~It is assumed that close to the Hopf bifurcation, the first 
Fourier mode of the $x$ variable is the dominant one. Equating like harmonics on 
both sides of the resulting equation, one obtains expressions for ~$F_0$, ~$F_1$,  
~$F_2$, ~etc.  In the absence of any external stimulus, the $F_i$'s vanish, 
enabling one to obtain explicit expressions for $x_i$s. Since the first Fourier 
mode ~$x_1$ is assumed to dominate near the bifurcation, we obtain :
\begin{eqnarray}
x_0  &\approx & p_4(C_0) + p_5(C_0){|x_1|}^2 \\
x_2  &\approx & - \frac{\lambda_2(C_0)}{\lambda_1(C_0)} {|x_1|}^2 \\
0  &\approx & \lambda_6(C_0)x_1 + \lambda_8(C_0){|x_1|}^2x_1
\end{eqnarray}
where ~~$p_4$ and ~$p_5$ are real and ~$\lambda_l$  ~(l= 1, 2, 6, 8) ~ 
complex, and they are all complicated functions of ~$C_0$, and of the expansion 
coefficients  ~$\zeta_{k, k_1,\dots, k_n}^{(n)}$, ~${\cal F}_{k, k_1,\dots, 
k_n}^{(n)}$, ~$G_{k, k_1,\dots, k_n}^{(n)}$, ~$H_{k, k_1,\dots, k_n}^{(n)}$ and 
~$M_{k, k_1,\dots, k_n}^{(n)}$. \\ 
The dependence of the stationary part ~$x_0$ on ~$|x_1|$ can be eliminated 
by an appropriate choice of ~$p_5(C_0)$ and ~$p_4(C_0)$. We work with the choice 
~$p_4 = p_5 = 0$. In the case of the hair bundle, where the $x$ variable 
corresponds to the hair bundle displacement, this choice of ~$p_4$ and $p_5$ 
would correspond to the elimination of the undesirable dependence of the 
constant part ~$x_0$ on the motion of the system. $p_4$ and ~$p_5$ in this 
biological example would depend upon the mechanical properties of the hair 
bundle --- however, we do not discuss this issue in our paper.  

\noindent From eqn.(A-6) we arrive at eqn.(6). 
In general, eqn.(6) has the structure 
\begin{equation}
{|x_1|}^2 \approx -\frac{a_0 + a_1C_0 + a_2C_0^2 + a_3C_0^3 + \dots}
{b_0 + b_1C_0 + b_2C_0^2 + b_3C_0^3 + \dots}
\end{equation}
where ~$a_i$s and ~$b_i$s depend upon the frequency.  Thus, writing eqn.(A-7) in the 
form of eqn.(7) implies that
\begin{equation}
C_c = C_0 {\Big\{ 1 + \Big[\frac{a_0 + a_1C_0 + a_2C_0^2 + a_3C_0^3 + \dots}
{b_0 + b_1C_0 + b_2C_0^2 + b_3C_0^3 + \dots}\Big] \Big\}}^{-1} .
\end{equation}

\subsection*{Appendix B : Calculation of noise averages}
\renewcommand{\theequation}{B-\arabic{equation}} 
\setcounter{equation}{0}
The mean values of functions ~$G(z, \bar z)$ are the noise averages 
\begin{equation}
\langle G(z, \bar z)\rangle = \int dz d{\bar z}G(z,\bar z)P(z, \bar z, t) 
\end{equation}
which may in general be calculated in the polar coordinate representation using 
the reduction given by eqn.(41). For instance, the mean value of ~$x(t)$ can be 
calculated using eqn.(2) and eqn.(16) 
\begin{equation}
\langle x(t)\rangle = \int_0^\infty rdr \int_0^{2\pi} r\cos \phi d\phi P(r,\phi,t) .
\end{equation}
From eqn.(41) and eqn.(42) we can calculate its asymptotic mean value 
\begin{equation}
\langle x(t)\rangle_{as} = \int_0^\infty dr \int_0^{2\pi}d\phi r^2\cos\phi 
{\cal P}(\phi |r,\lambda){\cal P}_{as}(r,\lambda,t) 
\end{equation}
In the adiabatic limit we can evaluate this using eqn.(45) for the forced system 
\begin{eqnarray}
\langle x(t)\rangle_{ad} &=& \frac{{\cal N}_F}{2\pi}\int_0^\infty dr \int_0^{2\pi}d\phi  
\delta(\phi -\Omega t -\alpha \pm 2n\pi) \cos\phi \nonumber\\
&\times& r^3 \exp{\Big\{ -\frac{2}{Q\epsilon}
\Big[ (l - \frac{\beta'C_c}{4\Delta^2})\frac{r^4}{4} 
- \beta\frac{r^2}{2} - F_0r \Big] \Big\}}
\end{eqnarray} 
Performing this integral leads to eqn.(52). \\

\subsection*{Acknowledgements} 
I am deeply indebted to Prof. Raymond Kapral for very helpful and productive 
discussions and useful comments, for informing me about references [28-31] and 
letting me have a copy of [31], and to Prof.Henry Tuckwell, Prof. Holger Kantz, 
Prof. J. Jost \& Dr. F. Atay for very helpful discussions. 
I am also very grateful to Prof. P. H\"anggi for a short discussion on 
stochastic resonance and Dr. W. Just for clarifications \& suggestions over 
email during the earlier stages of this work. My thanks to Drs. Markus Porto, 
V.Shatokhin, N.Vitanov \&  R.Klages for discussions at various times, 
to Dr.Ellen Lumpkin for very useful information and a discussion on the 
hair bundle and to Dr.B.Ashok for very helpful comments \& suggestions 
on the manuscript. Finally, I would like to acknowledge 
hospitality of the Max-Planck-Institut f\"ur Physik komplexer Systeme, Dresden, 
where the major part of this project was completed.

\newpage 
\subsection*{References} 
\begin{enumerate}
\item L.Arnold, {\em Random Dynamical Systems}, Springer-Verlag, (1998). 
\item J.L.Cabrera \& J.G.Milton,{\em Phys.Rev.Lett.}, {\bf 89}, 158702 (2002).  
\item D.P.Corey \& A.J.Hudspeth, {\em J.Neurosci.}, {\bf 3}, 962 (1983).
\item J.Howard \& A.J.Hudspeth, PNAS, {\bf 84} 3064 (1987); \\
J.Howard \& A.J.Hudspeth, {\em Neuron}, {\bf 1}, 189 (1988).
\item M.A.Ruggero, {\em Curr.Opin.Neurobiol.}, {\bf 2}, 449 (1992).
\item R.A.Eatock, D.P.Corey \& A.J.Hudspeth, {\em J.Neurosci.}, {\bf 7}, 2821 
(1987); \\
A.J.Hudspeth, {\em Nature}, {\bf 341}, 397 (1989);\\
J.A.Assad, N.Hacohen \& D.P.Corey, {\em PNAS}, {\bf 86} 2918 (1989);\\ 
E.A.Lumpkin \& A.J.Hudspeth, {\em PNAS}, {\bf 92} 10297 (1995);\\
E.A.Lumpkin \& A.J.Hudspeth, {\em J.Neurosci.}, {\bf 18}, 6300 (1998);\\
P.Dallos, {\em J.Neurosci.}, {\bf 12}, 4575 (1992);\\ 
J.A.Assad \& D.P.Corey, {\em J.Neurosci.}, {\bf 12}, 3291 (1992);\\ 
A.J.Hudspeth \& P.G.Gillespie, {\em Neuron}, {\bf 12}, 1 (1994);\\
A.C.Crawford \& R.Fettiplace, {\em J.Physiol.}, {\bf 312}, 377 (1981);\\
G.A.Manley, {\em J.Neurophysiol.}, {\bf 86}, 541 (2001);\\  
A.J.Hudspeth, Y.Choe, A.D.Mehta \& P.Martin, {\em PNAS}, {\bf 97}, 11765 (2000).\\ 
F.Jaramillo, V.S.Markin \& A.J.Hudspeth, {\em Nature}, {\bf 364}, 527 (1993).
\item T.Gold, {\em Proc.R.Soc.B}, {\bf 135}, 492 (1948).
\item P.M.Zurek, {\em J.Acoust.Soc.Am.}, {\bf 69}, 514 (1981);\\
C.K\"oppl, in {\em Advances in Hearing Research}(ed. G.A.Manley, C.K\"oppl, 
H.Fastl \& H.Oeckinghaus), pp.200-209, World Scientific, Singapore (1995).
\item M.C.G\"opfert \& D.Robert, {\em Proc.R.Soc.Lond.}, {\bf B 268}, 333 (2000). 
\item Y.Choe, M.O.Magnasco \& A.J.Hudspeth, {\em Proc.Natl.Acad.Sci.USA}, {\bf 95}, 
15321 (1998).
\item S.Camalet, T.Duke, F.J\"ulicher \& J.Prost, {\em PNAS}, {\bf 97}, 
3183 (2000)). 
\item V.M.Eguiluz, M.Ospeck, Y.Choe, A.J.Hudspeth \& M.O.Magnasco, 
{\em Phys.Rev.Lett.}, {\bf 84}, 5232 (2000);\\
M.Ospeck, V.M.Eguiluz \& M.O.Magnasco, Biophys.J, {\bf 80}, 2597 (2001). 
\item A.Vilfan \& T.Duke, {\em Biophys.J}, {\bf 85}, 191 (2003).
\item F.J\"ulicher, D.Andor \& T.Duke, {\em PNAS}, {\bf 98}, 9080 (2001)). 
\item L.Moreau \& E.Sontag, {\em Phys.Rev.}, {\bf E 68}, 020901(R) (2003);\\ 
L.Moreau, E.Sontag \& M.Arcak, {\em Syst.Control Lett.}, {\bf 50}, 229 (2003).
\item F.Jaramillo \& K.Wiesenfeld, {\em Nature Neuroscience}, {\bf 1}, 384 (1998).
\item J.K.Douglas, L.Wilkens, E.Pantazelou \& F.Moss, {\em Nature}, 
{\bf 365}, 337 (1993).   
\item S.M.Bezrukov \& I.Vodyanoy, {\em Nature}, {\bf 378}, 362 (1995).
\item W.Denk \& W.W.Webb, {\em Hear.Res.}, {\bf 60}, 89 (1992). 
\item K.Wiesenfeld \& F.Moss, {\em Nature}, {\bf 373}, 33 (1995).
\item P.Jung \& K.Wiesenfeld, {\em Nature}, {\bf 385}, 291 (1997). 
\item R.Benzi, A.Sutera \& A.Vulpiani, {\em J.Phys.A} {\bf 14}, L453 (1981);\\
C.Nicolis \& G.Nicolis, {\em Tellus}, {\bf 33}, 275 (1981);\\
C.Nicolis, {\em Tellus}, {\bf 34}, 1 (1982);\\
R.Benzi, G.Parisi, A.Sutera \& A.Vulpiani, {\em Tellus}, {\bf 34}, 10 (1982).
\item B.McNamara \& K.Wiesenfeld, {\em Phys.Rev.} {\bf A 39}, 4854 (1989).
\item P.Jung \& P.H\"anggi, {\em Europhys.Lett.} {\bf 8}, 505 (1989);\\
P.Jung \& P.H\"anggi, {\em Phys.Rev.} {\bf A 41}, 2977 (1990);\\
L.Gammaitoni, P.H\"anggi, P.Jung \& F.Marchesoni, {\em Rev.Mod.Phys.} {\bf 70}, 223 
(1998).
\item P.Jung \& P.H\"anggi, {\em Phys.Rev.} {\bf A 44}, 8032 (1991). 
\item V.S.Anishchenko, A.B.Neiman, F.Moss \& L.Schimansky-Geier, 
{\em Uspekhi Fizicheskih Nauk} {\bf 169}, 7 (1999); 
{\em Sov.Phys.Usp.} {\bf 42}, 7 (1999).
\item J.Guckenheimer \& P.Holmes, {\em Nonlinear oscillations, dynamical systems 
\& bifurcations of vector fields}, Springer-Verlag, (1983).
\item J.M.Gambaudo, {\em J.Diff.Equations}, {\bf 57}, 172 (1985).
\item C.Elphick, G.Iooss \& E.Tirapegui, {\em Phys.Lett.}, {\bf 120A}, 459 (1987).
\item C.Hemming \& R.Kapral, {\em Faraday Discuss.}, {\bf 120}, 371 (2001).
\item C.Hemming, PhD thesis, University of Toronto (December 2002).
\item M.Schumaker, {\em Phys.Lett.}, {\bf 122A}, 317 (1987).
\item C.Van den Broeck, M.Malek Mansour \& F.Baras, {\em J.Stat.Phys}, {\bf 28}, 
557 (1982);\\
F.Baras,M.Malek Mansour \& C.Van den Broeck, {\em J.Stat.Phys}, {\bf 28}, 577 (1982).
\item N.G.Van Kampen, {\em Stochastic processes in physics and chemistry}, 
(revised edition), Elsevier, (1992);\\ 
C.W.Gardiner, {\em Handbook of Stochastic Methods for Physics, Chemistry \&
 the Natural Sciences} (2nd edition), Springer Verlag (1985). 
\item P.H.Coullet, C.Elphick \& E.Tirapegui, {\em Phys.Lett.}, {\bf 111A}, 277 (1985).
\item E.Knobloch \& K.A.Wiesenfeld, {\em J.Stat.Phys}, {\bf 33}, 611 (1983).
\item S.Siegmund, {\em J.Diff.Eq.}, {\bf 178}, 541 (2002).
\item N.Sri Namachchivaya \& S.T.Ariaratnam, {\em SIAM J.Appl.Math.}, 
{\bf 47}, 15 (1987).
\item J.Langa, J.C.Robinson \& A.Suarez, {\em Nonlinearity}, {\bf 15}, 887 (2002);\\
A.M.Mancho, D.Small, S.Wiggins \& K.Ide, {\em Physica} {\bf D 182}, 188 (2003).
\item N.Sri Namachchivaya, {\em Int.J.Nonlinear Mech},{\bf 26}, 931 (1991). 
\item I.S.Gradshteyn \& I.M.Ryzhik, {\em Table of Integrals, Series, and Products} 
(5th edition), Academic Press, (1994). 
\item K.Wiesenfeld, {\em J.Stat.Phys}, {\bf 38}, 1071 (1985). 
\item A.Neiman, P.I.Saparin \& L.Stone, {\em Phys.Rev.} {\bf E 56}, 270 (1997). 
\item L.P.Shilnikov, A.L.Shilnikov, D.V.Turaev \& L.O.Chua, {\em Methods 
of Qualitative Theory in Nonlinear Dynamics, Part 1}, World Scientific, 
Singapore, (1998). 
\item L.Omberg, K.Dolan, A.Neiman \& F.Moss, {\em Phys.Rev.}, {\bf E 61}, 
4848 (2000). 
\item G.Zweig, in {\em  Biophysics of the cochlea} (ed. A.W.Gummer), pp 315-329, 
World Scientific, Singapore (2002).  
\item R.Nobili, A.Vetesnik, L.Turicchia \& F.Mammano, {\em J Assoc Res Otolaryngol.}, 
{\bf 4}, 478 (2003). 

\end{enumerate}

\newpage

{\includegraphics[height=6cm,width=8.5cm]{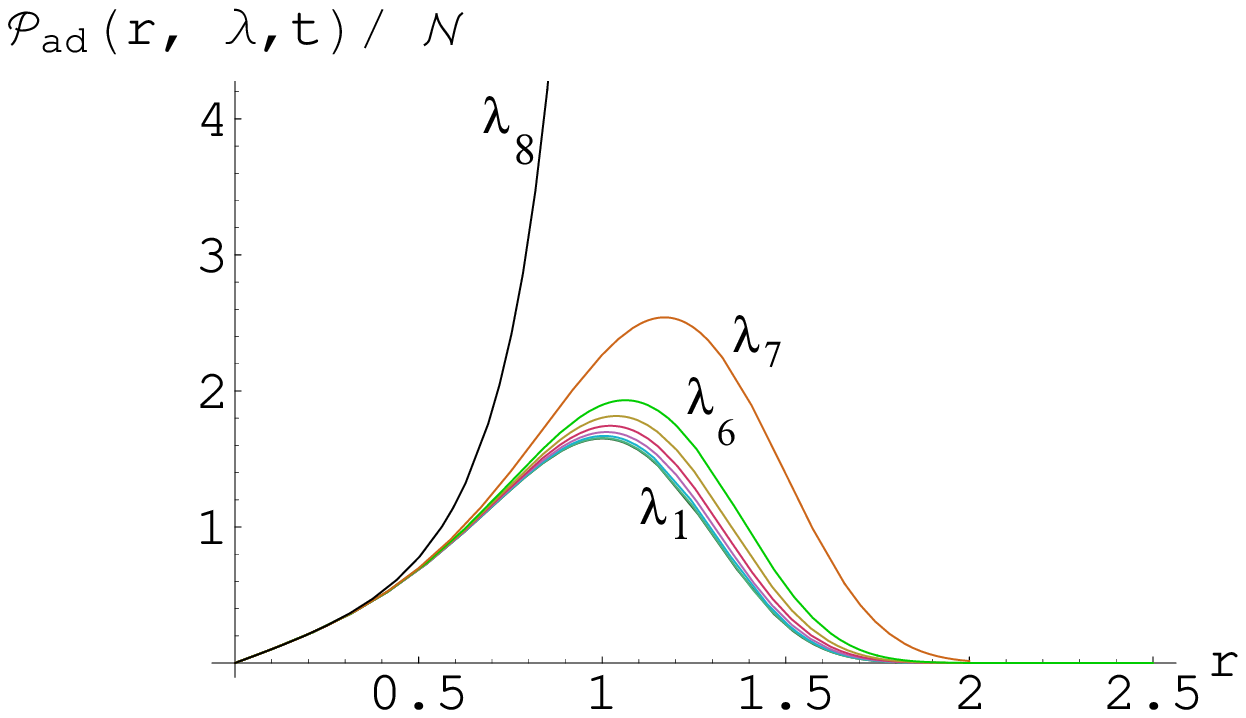}}
\vspace*{.5in}\\ 
{\bf Fig.1.~~~Effect of Self-tuning: ~Un-normalised probability distribution 
~$\frac{{\cal P}_{ad}(r,\lambda,t)}{\cal N}$  as a function of the radial 
variable  ~$r$ ~ for increasing values of ~$\lambda$.} ~
($l=2.6\times 10^{-3}; ~\beta=1.5\times 10^{-3}; 
~\frac{\beta'C_c}{4\Delta^2}=1.2\times 10^{-3}; ~Q\epsilon=1.0\times 10^{-3}; 
~\lambda_1=0.05, ~\lambda_2=0.2, ~\lambda_3=0.3, ~\lambda_4=0.4, 
~\lambda_5=0.5, ~\lambda_6=0.6, ~\lambda_7=0.8,  ~\lambda_8=1.2$).\\

{\includegraphics[height=6cm,width=8.5cm]{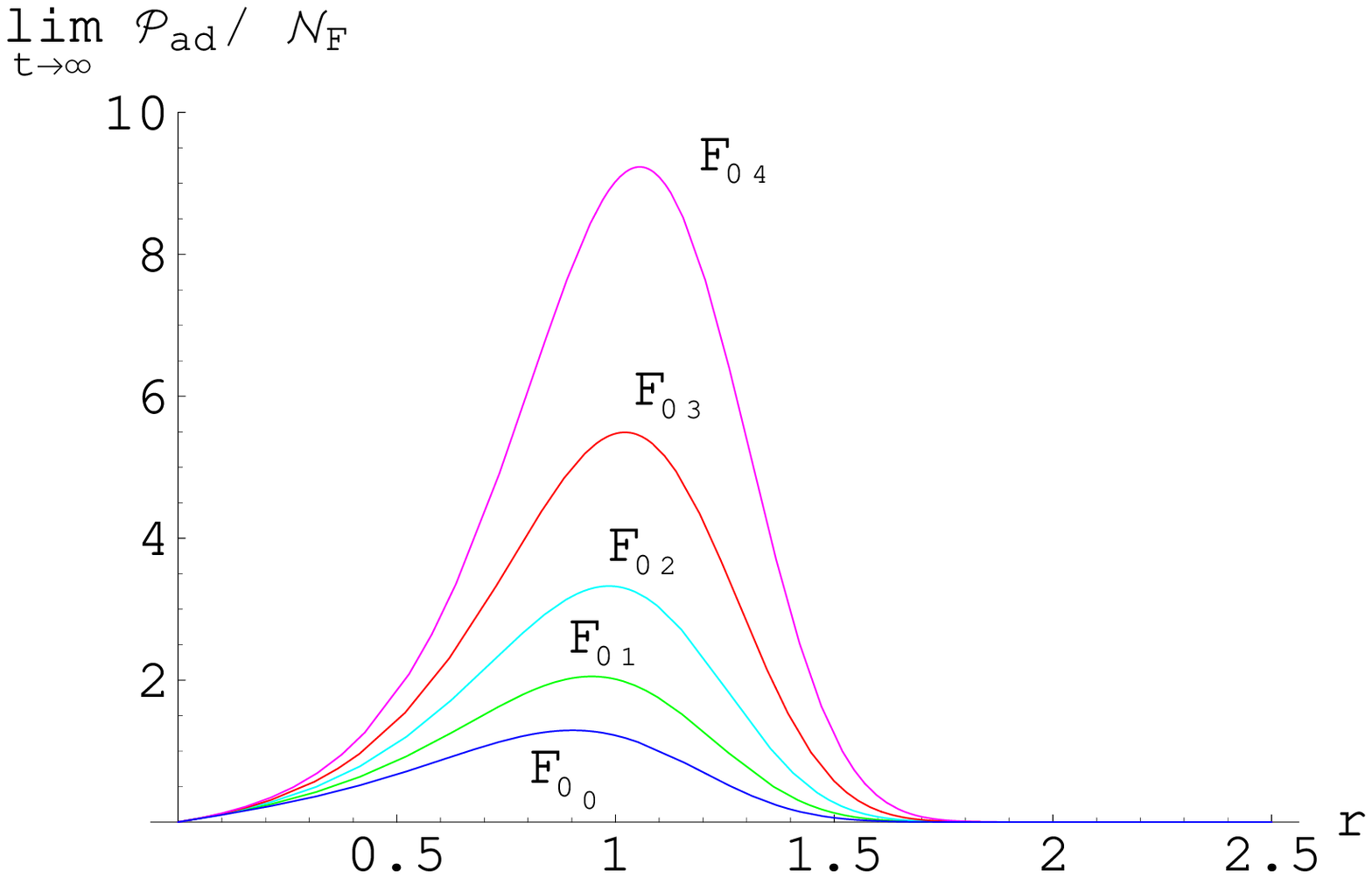}}
\vspace*{.5in}\\
{\bf Fig.2.~~~Probability distribution for a forced Hopf oscillator for 
different values of the external forcing $F_0$}. ~( $l=3.8\times 10^{-3}; 
~\beta=1.5\times 10^{-3}; ~\frac{\beta'C_c}{4\Delta^2}=1.2\times 10^{-3}; 
~Q\epsilon=1.0\times 10^{-3}; ~{F_0}_0=0.0; ~{F_0}_1= 0.5\times 10^{-3}; 
~{F_0}_2= 1.0\times 10^{-3}; ~{F_0}_3= 1.5\times 10^{-3}; 
~{F_0}_4= 2.0\times 10^{-3}$). \\ 

\noindent
{\includegraphics[height=6cm,width=8.5cm]{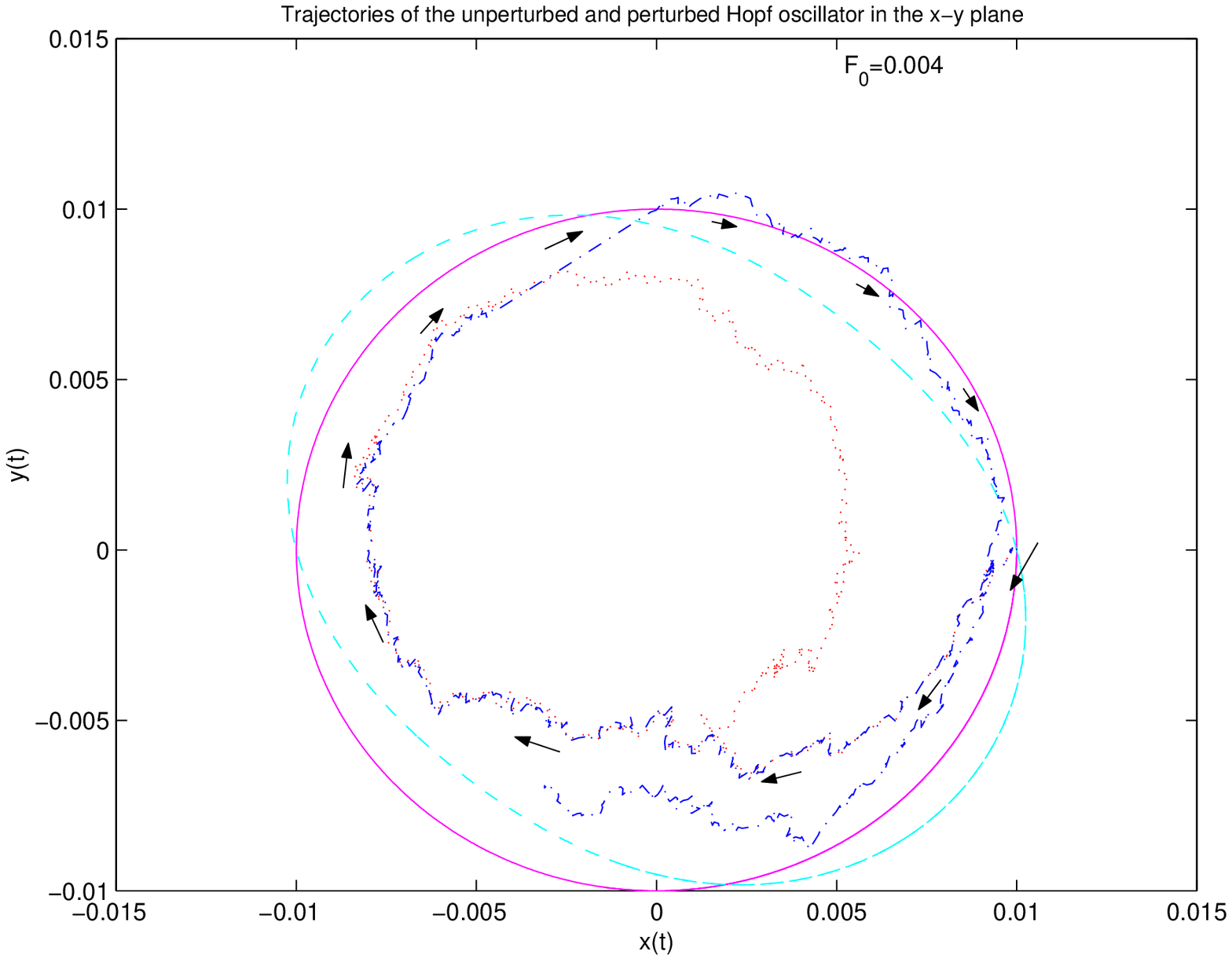}}
\hspace*{0.5cm}
{\includegraphics[height=6cm,width=8.5cm]{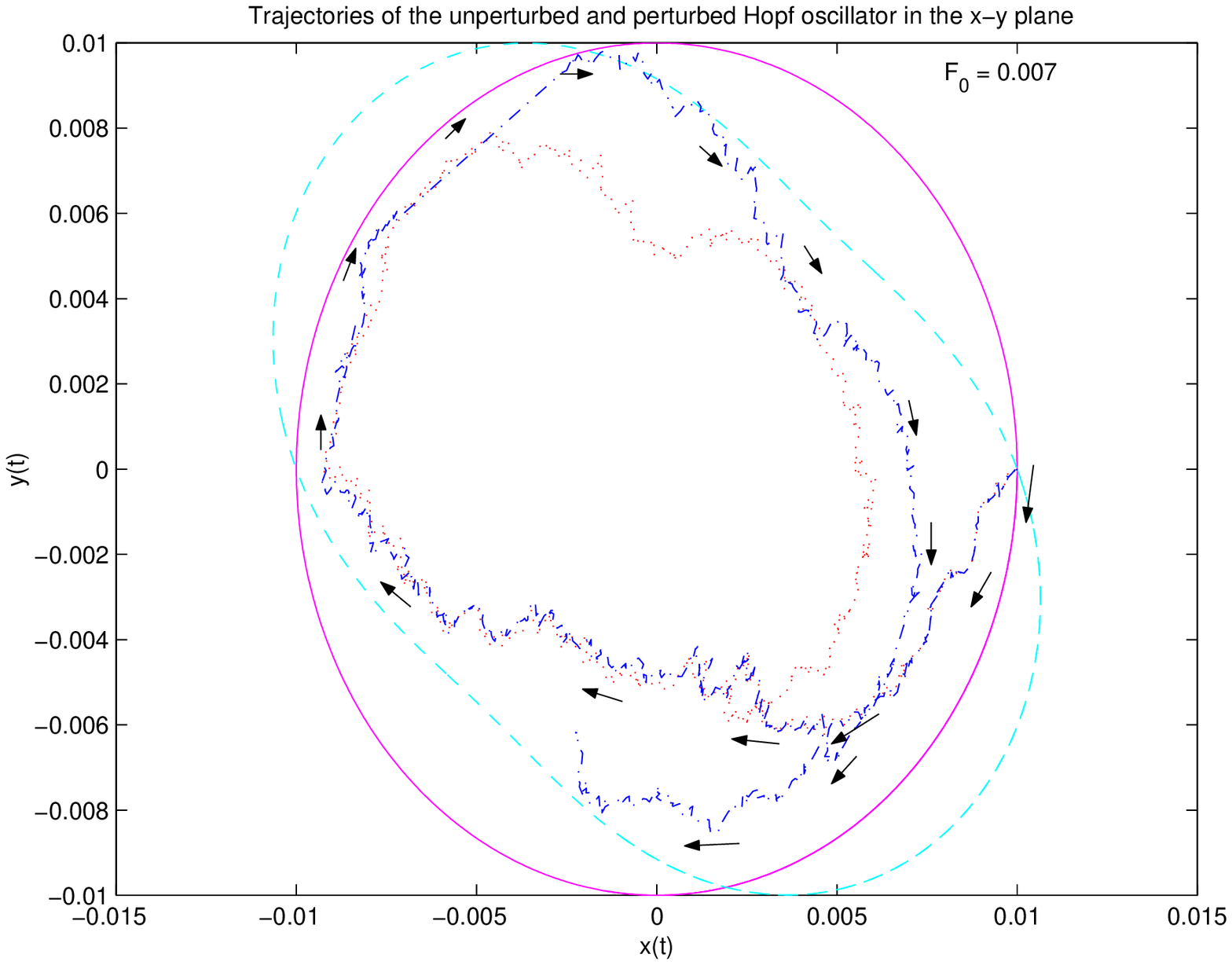}}\\
\vspace*{0.2cm}\\
\noindent
3A. \hspace{8.5cm} 3B.\\
\noindent
\vspace*{.5in}\\ 
{\bf Fig.3.~~~Trajectories in the x-y plane of the unperturbed and 
perturbed Hopf 
oscillators for two different amplitudes of periodic forcing }: ~the solid 
closed curve is the trajectory of the unforced 
deterministic Hopf oscillator without feedback control; ~the dashed trajectory 
is of the forced deterministic Hopf oscillator without feedback control; ~
the dotted one is of the noisy forced Hopf oscillator without feedback 
control; ~and the dash-dotted trajectory shows the effect of the
feedback control on the noisy forced Hopf oscillator. ($l=0.1; ~\beta=0.0; 
~\frac{C_c}{4\Delta^2}= 0.3; ~d=0.4; ~\alpha=0.2; ~\alpha_0=0.2; 
~\omega=1.6; ~\Omega=1.6; ~\beta'=0.00001; ~\epsilon=0.0009$). 
({\bf Fig.3A.}~$F_0=0.004$, ~~{\bf Fig.3B.}~$F_0=0.007. $).\\

{\includegraphics[height=6cm,width=8.5cm]{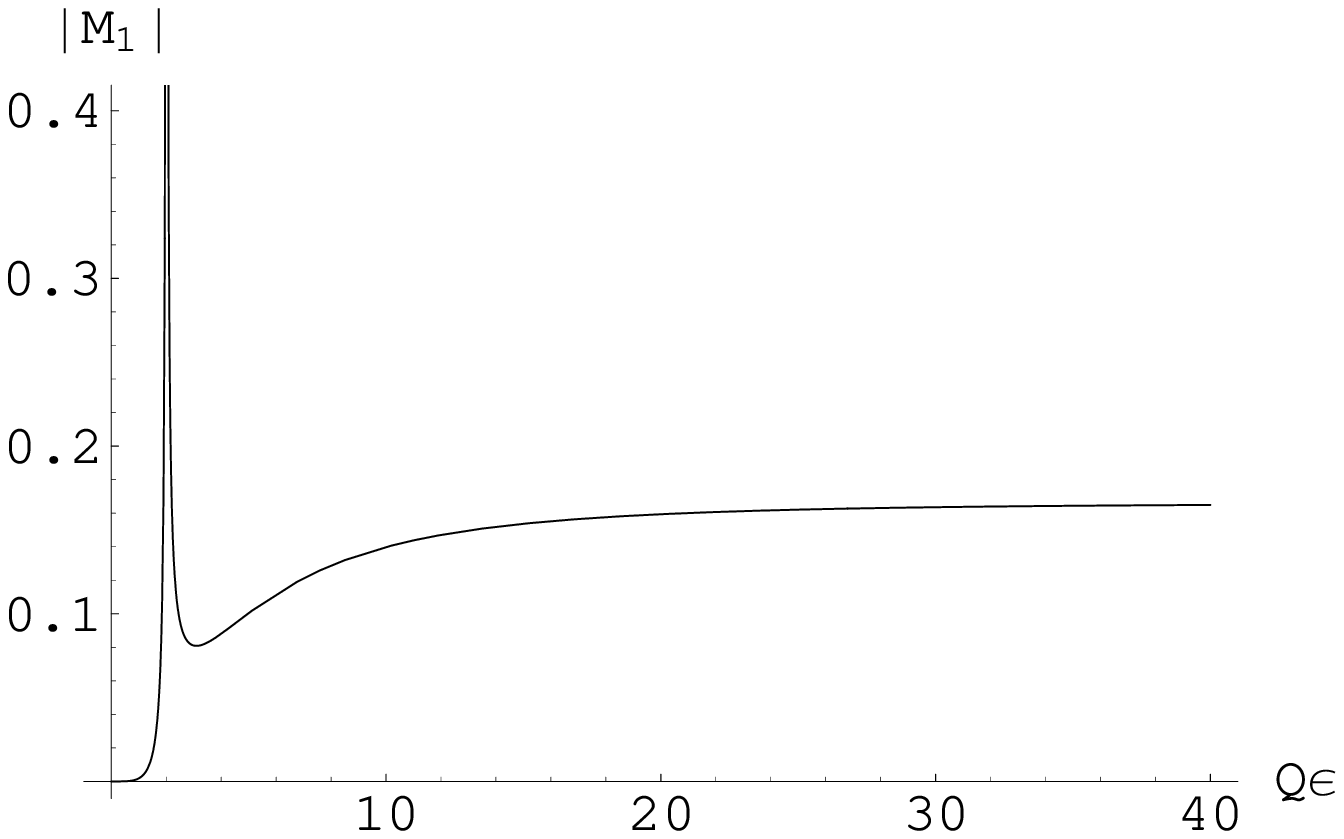}
\vspace*{.5in}\\
{\bf Fig.4.~~~Response ~$|M_1|$ of a resonantly forced self-tuned 
Hopf-oscillator in the presence of additive noise  ~$Q\epsilon$ 
(for ~$F_0=1.5\times 10^{-3}, ~\beta=0.1, ~l-\frac{\beta'C_c}
{4\Delta^2}=1.6\times 10^{-3} $ ). }\\

{\includegraphics[height=6cm,width=8.5cm]{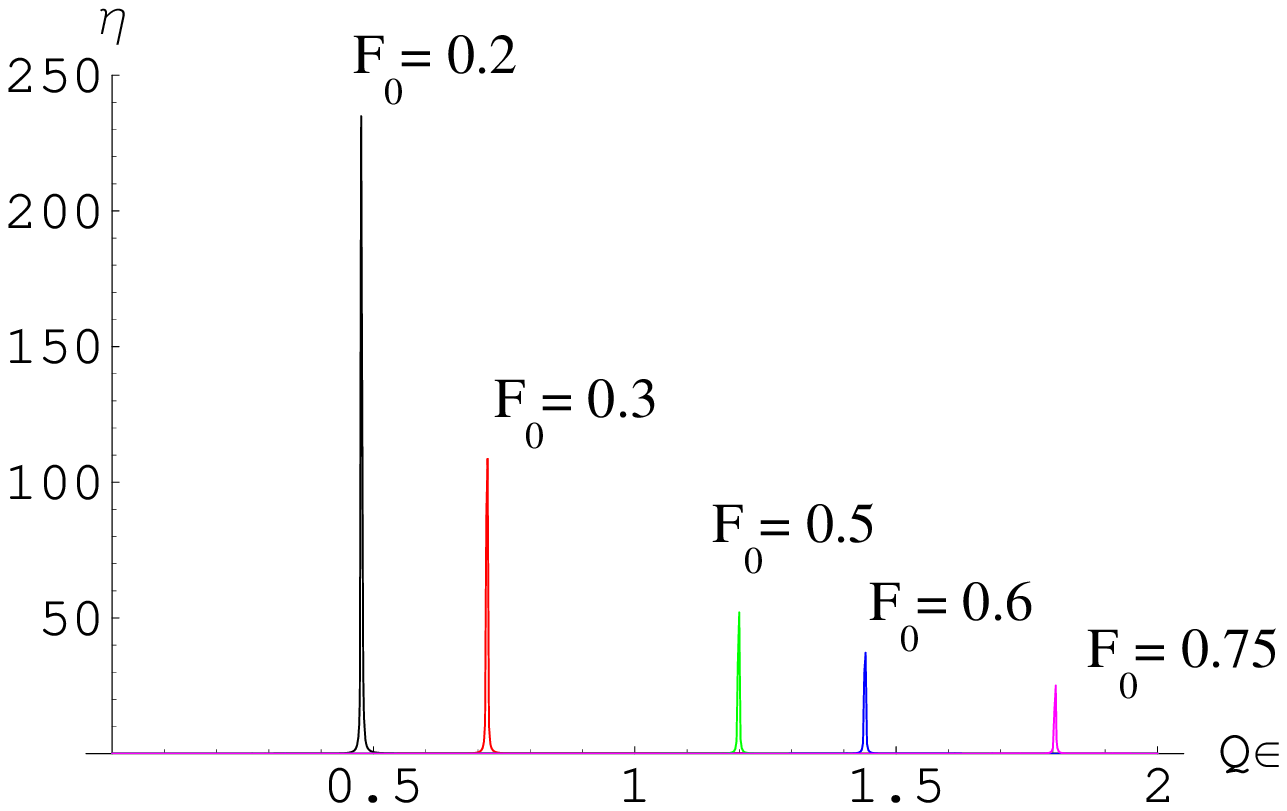}}
\vspace*{.5in}\\
{\bf Fig.5.~~~Spectral power amplification ~$\eta$ of a resonantly forced 
self-tuned Hopf oscillator as a function of the noise strength ~$Q\epsilon$ 
~(for $\beta=0.1, ~l-\frac{\beta'C_c}{4\Delta^2}=0.14\times 10^{-3} $).}\\

{\includegraphics[height=6cm,width=8.5cm]{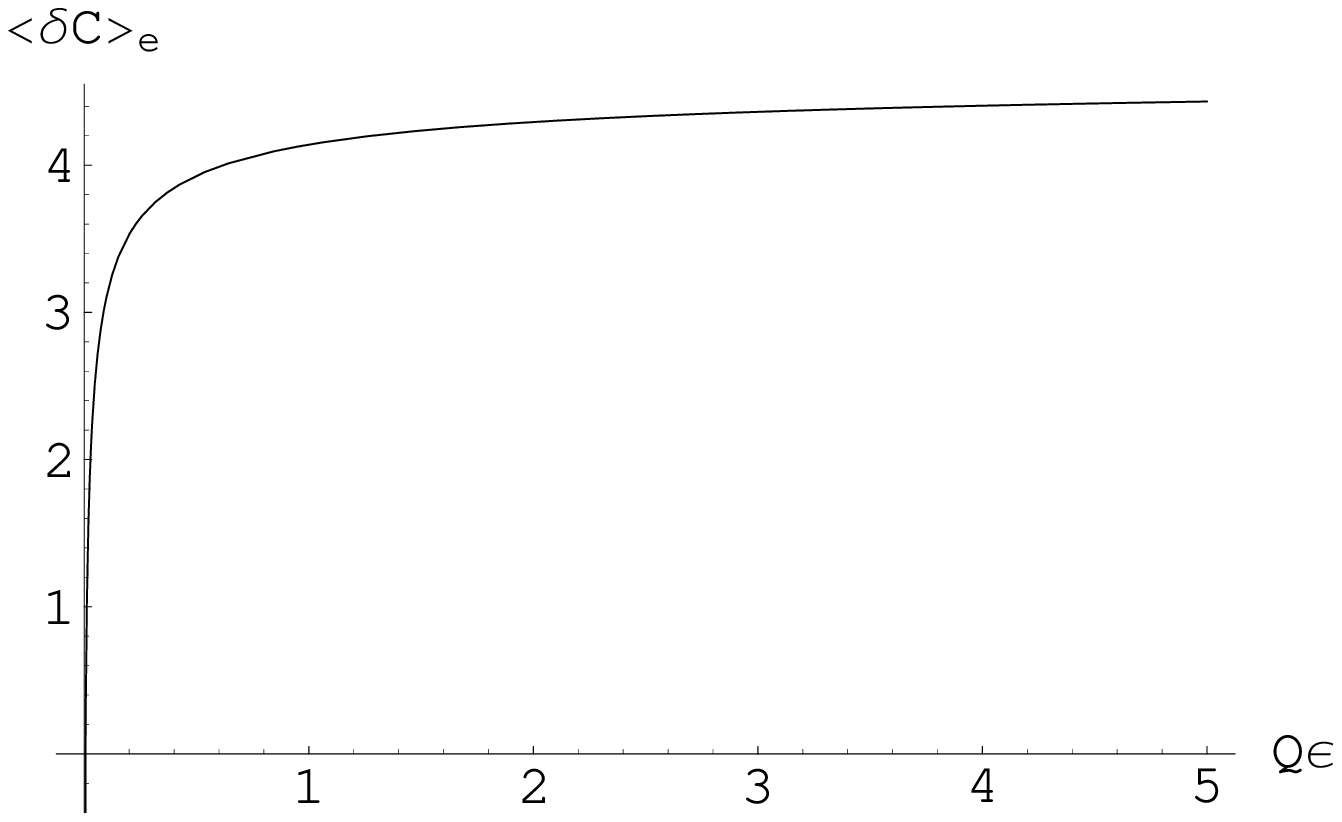}}
\vspace*{.5in}\\
{\bf Fig.6.~~~Self-tuning in a forced Hopf oscillator subject to additive 
noise: ~ Distance ~$\langle\delta C\rangle_e$ ~from the 
bifurcation  as a function of the noise strength ~$Q\epsilon$ 
(for ~$\frac{F_0}{\beta}=0.02$).}\\

{\includegraphics[height=6cm,width=8.5cm]{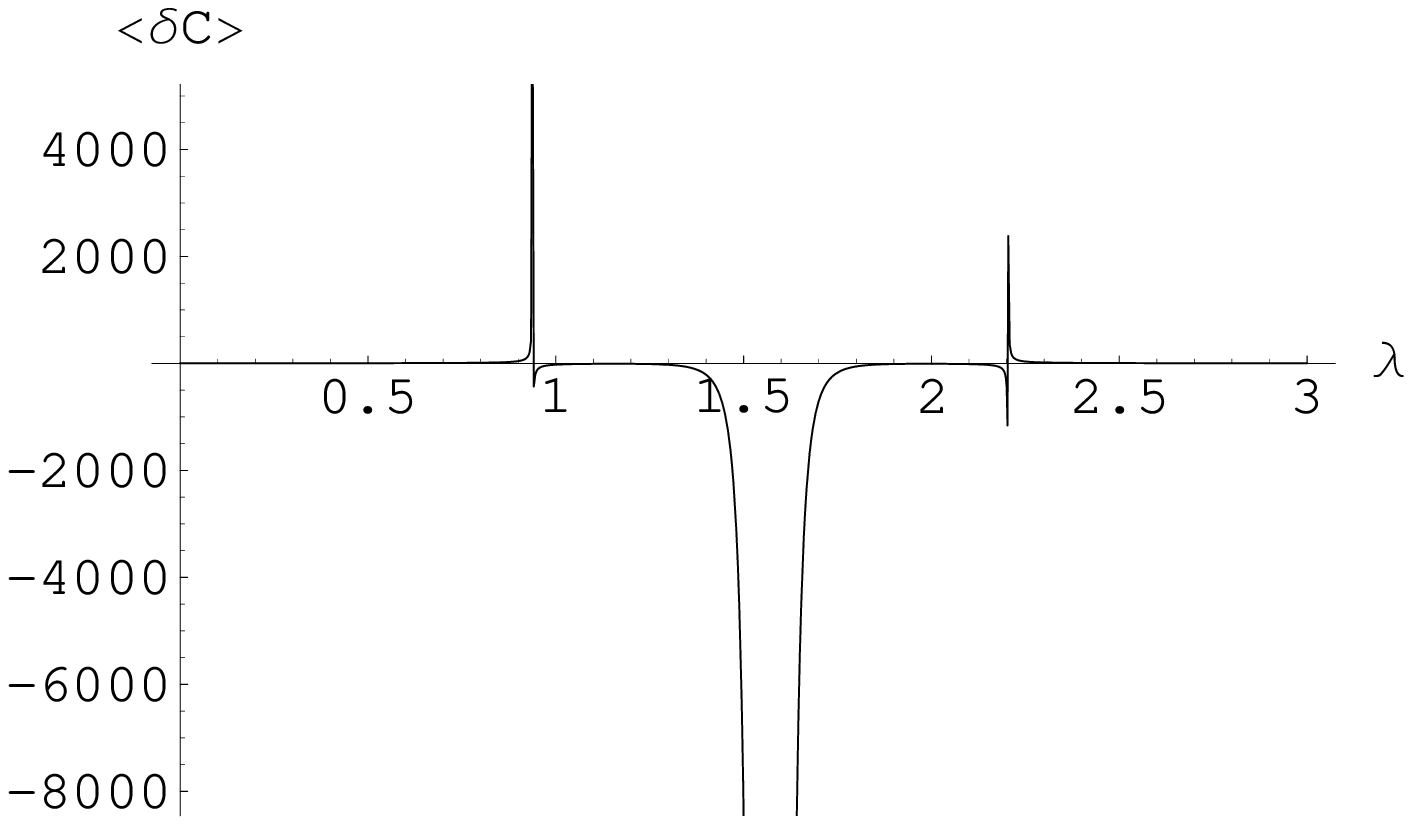}}
\vspace*{.5in}\\
{\bf Fig.7.~~~Self-tuning in a noisy Hopf oscillator in the absence 
of external force: ~ Distance ~$\langle\delta C\rangle$ ~from the 
bifurcation is periodically modulated for fixed noise strength, 
spiking at specific frequencies  ~($\beta^2=1.5\times 10^{-6}, 
~l=2.0\times 10^{-3}, 
~\frac{\beta'C_c}{8\Delta^2}=0.7\times 10^{-3}, ~Q\epsilon =1.0\times 10^{-3}$).}\\ 

\end{document}